\documentclass[aps,prl,preprint,superscriptaddress]{revtex4-1}

\usepackage{graphics,graphicx}
\usepackage{epstopdf}

\begin{document}


\title{Fluctuating Nonlinear Spring Model of Mechanical Deformation of Biological Particles}


\author{Olga Kononova}
\affiliation{Department of Chemistry, University of Massachusetts, Lowell, MA 01854, USA}
\affiliation{Moscow Institute of Physics and Technology, Moscow Region, 141700, Russia}

\author{Joost Snijder}
\affiliation{Natuur- en Sterrenkunde and LaserLab, Vrije Universiteit, 1081 HV Amsterdam, The Netherlands}

\author{Kenneth A. Marx}
\affiliation{Department of Chemistry, University of Massachusetts, Lowell, MA 01854, USA}

\author{Gijs J. L. Wuite}
\affiliation{Natuur- en Sterrenkunde and LaserLab, Vrije Universiteit, 1081 HV Amsterdam, The Netherlands}

\author{Wouter H. Roos}
\email[Corresponding author: ]{wroos@few.vu.nl}
\affiliation{Natuur- en Sterrenkunde and LaserLab, Vrije Universiteit, 1081 HV Amsterdam, The Netherlands}

\author{Valeri Barsegov}
\email[Corresponding author: ]{Valeri\_Barsegov@uml.edu}
\affiliation{Department of Chemistry, University of Massachusetts, Lowell, MA 01854, USA}
\affiliation{Moscow Institute of Physics and Technology, Moscow Region, 141700, Russia}


\date{\today}

\begin{abstract}
We present a new theory for modeling forced indentation spectral lineshapes of biological particles, which considers non-linear Hertzian deformation due to an indenter-particle physical contact and bending deformations of curved beams modeling the particle structure. The bending of beams beyond the critical point triggers the particle dynamic transition to the collapsed state, an extreme event leading to the catastrophic force drop as observed in the force ($F$)-deformation ($X$) spectra. The theory interprets fine features of the spectra: the slope of the FX curves and the position of force-peak signal, in terms of mechanical characteristics --- the Young's moduli for Hertzian and bending deformations $E_H$ and $E_b$, and the probability distribution of the maximum strength with the strength of the strongest beam $F_b^*$ and the beams' failure rate $m$. The theory is applied to successfully characterize the $FX$ curves for spherical virus particles --- CCMV, TrV, and AdV.
\end{abstract}


\maketitle

\begin{figure}
\includegraphics[width=0.9\textwidth]{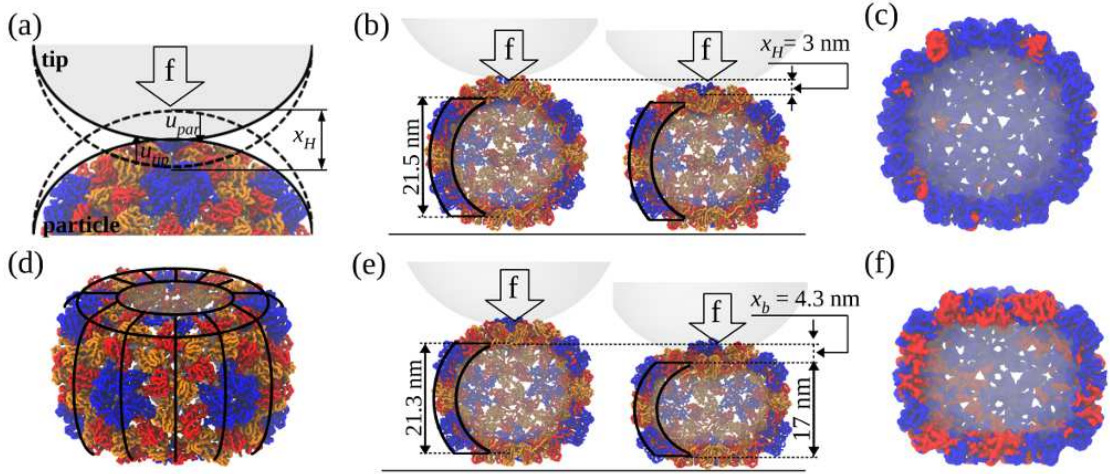}%
\caption{\label{fig:fig1} (color) Types of mechanical excitations exemplified using the CCMV shell. (a)-(c) Hertzian deformation $x_H$ with normal displacements $u_{tip}$ and $u_{par}$ (scheme on (a)) under the influence of compressive force (vertical arrow). Dashed contour lines show the tip and particle in their undeformed states. Structures in (b) --- the native (left) and partially deformed (right) states show the amplitude of $x_H$$\approx$$3$ nm. (c) CCMV shell profile showing parts of the structure with high potential energy ($>3$ kcal/mol per residue; red) and low potential energy (blue). (d)-(f) Beam-bending deformation. The side portion of the structure is partitioned into curved vertical beams (top-side view on (d)). Structures in (e) --- the partially deformed (left) and pre-collapse (right) states reveal the amplitude of $x_b$$\approx$$4.3$ nm. (f) CCMV shell profile under Hertzian and beam-bending deformations showing the potential energy distribution.}
\end{figure}

\begin{figure}
\includegraphics[width=0.9\textwidth]{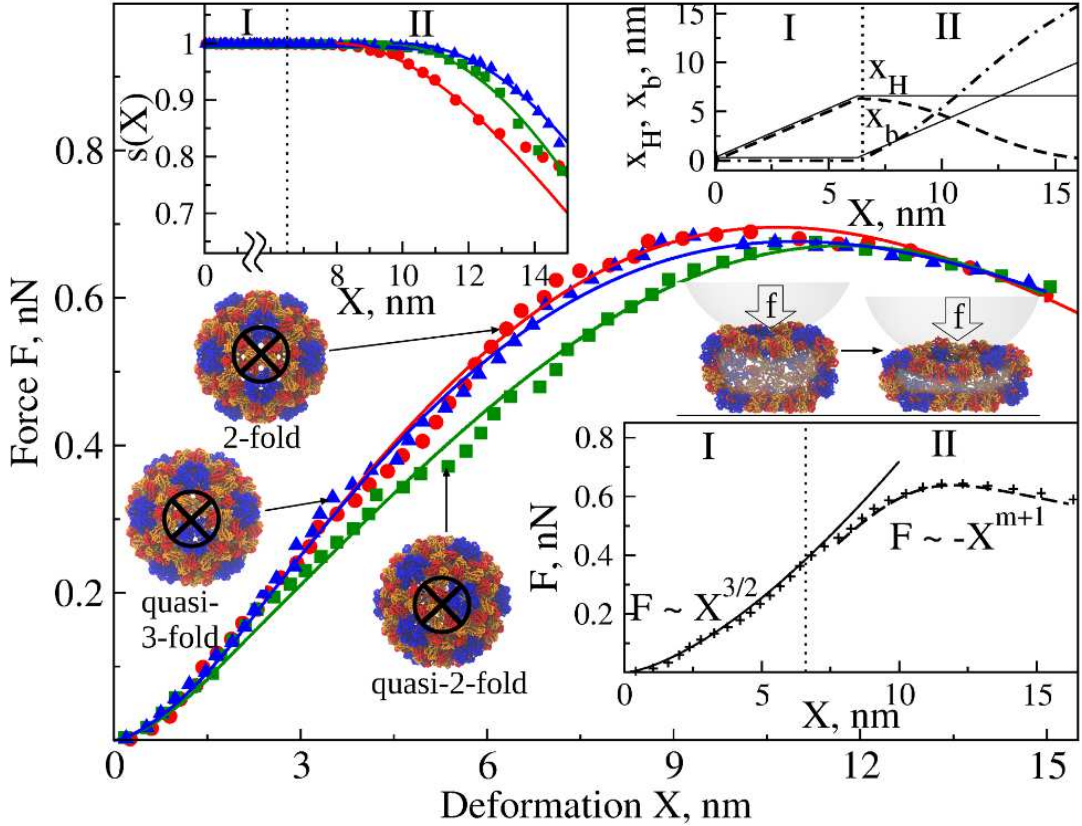}%
\caption{\label{fig:fig2} (color) Forced deformation of CCMV shell \textit{in silico}: the average $FX$ spectra (data points), obtained from the simulations of nanoindentation ($\nu_f$$=$$1.0$ $\mu$m/s, $R_{tip}$$=$$20$ nm, and $\kappa$$=$$0.05$ N/m; Fig. S4(a)) along the 2-fold (red), quasi-3-fold (blue), and quasi-2-fold symmetry axes (green), and theoretical $FX$ curves (solid lines). The circled bolded X shows the locations of force application. We used MD simulations accelerated on GPUs \cite{ZhmurovProt10,ZhmurovJPC11} and nanoindentations \textit{in silico} of the CCMV capsid \cite{KononovaBJ13} (Figs. S1-S3). Structures under the force maxima depict the capsid transitioning from the state before the collapse (left) to the collapsed state (right). The top insets: $\chi$-based estimation of $s(X)$ (left) and dynamics of $x_H$ and $x_b$ vs. $X$ (right) in the Hertzian regime I and in the transition regime II (dashed and dashed-dotted lines are for the exact method, and solid lines are for the approximate piece-wise method of parameter estimation). The bottom inset: schematic for piece-wise spectrum modeling; regime I --- $X$$\approx$$x_H$, and $F(X)$$\approx$$F_H$; regime II --- $X$$=$$x_H+x_b$ and $F(X)$$=$$F_H+F_b$.}
\end{figure}

\begin{figure}
\includegraphics[width=0.9\textwidth]{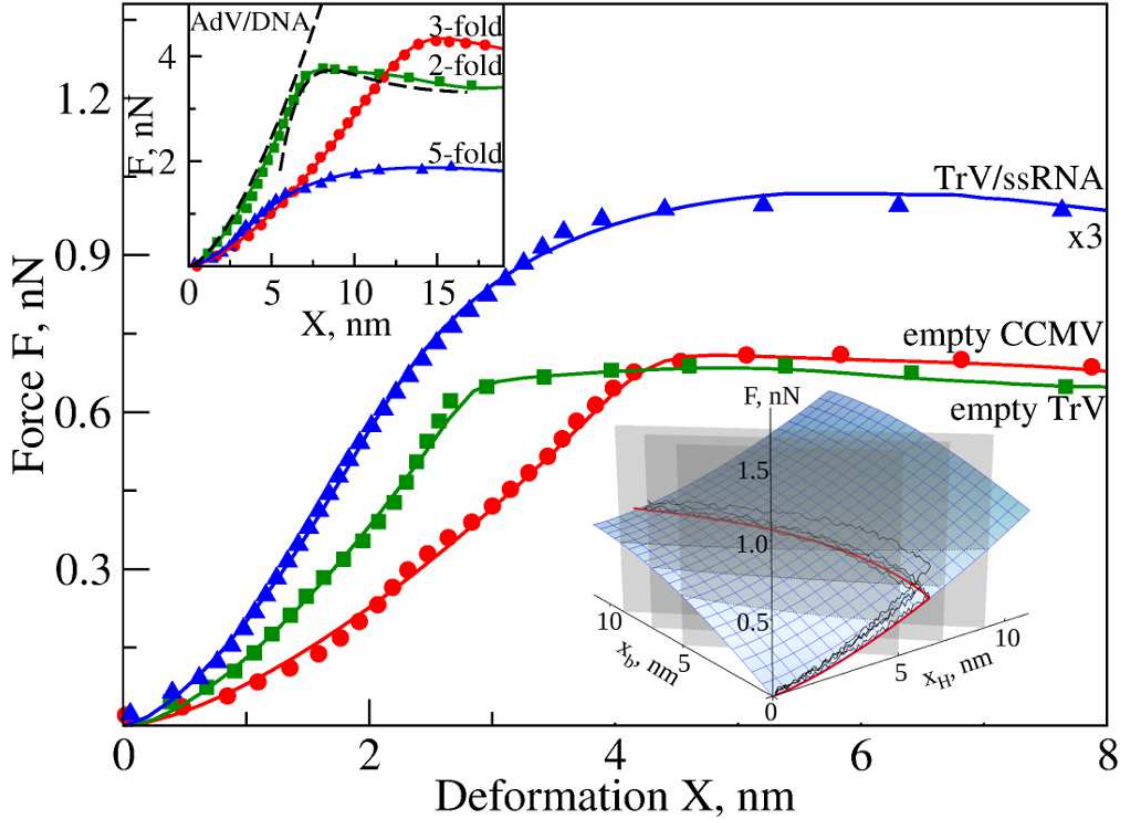}
\caption{\label{fig:fig3} (color) AFM-based forced deformation of empty CCMV and TrV shell, TrV with ssRNA, and AdV with dsDNA (top inset which also shows a schematic for piece-wise modeling); see Figs. S4(b), S5(a) - (c). For empty CCMV, $\nu_f$$=$$0.06$ $\mu$m/s, $R_{tip}$$=$$20$ nm, and $\kappa$$=$$0.05$ N/m. For empty TrV, $\nu_f$$=$$0.06$ $\mu$m/s, $R_{tip}$$=$$15$ nm, and $\kappa$$=$$0.056$ N/m. For TrV with ssRNA, $\nu_f$$=$$0.06$ $\mu$m/s, $R_{tip}$$=$$15$ nm, and $\kappa$$=$$0.1$ N/m. For full AdV with dsDNA, $\nu_f$$=$$0.055$ $\mu$m/s, $R_{tip}$$=$$15$ nm, and $\kappa$$=$$0.0524$ N/m. The average experimental spectra (data points) are compared with theoretical $FX$ (solid) curves. The bottom inset shows the 2D surface  (Eq.(\ref{eq:final})) constructed with the model parameters from CCMV indentation \textit{in silico} along the 2-fold symmetry axis (Fig.~\ref{fig:fig2}; Table I). The red curve is the equilibrium average curve $F_{eq}X_{eq}$ with the points formed by the intersection of $F(x_H,x_b)$ with line $x_b$$=$$X$$-$$x_H$ (shown for $X$$= 5,7$, and $9$ nm). The noisy black curves are particular realizations of the stochastic $FX$ path.}
\end{figure}

Mechanical testing has become the main tool to probe the physico-chemical and materials properties of the protein shells of plant and animal viruses, and bacteriophages \cite{RoosNatPhys10}. A variety of viruses have been explored by profiling the indentation force $F$ as a function of particle deformation $X$ ($FX$ curve), including bacteriophages $\Phi 29$ and $HK97$ \cite{IvanovskaPNAS04,HernandoPerezNano14,RoosPNAS12}, the human viruses Noro Virus, Hepatitis B Virus, Human Immuno Deficiency Virus (HIV), Adenovirus (AdV) and Herpes Simplex Virus \cite{KolBJ07,PerezBernaJBC12,RoosPNAS09,SnijderJV13} and other eukaryotic cell infecting viruses like Minute Virus of Mice, Triatoma Virus (TrV) and plant viruses Cowpea Chlorotic Mottle Virus (CCMV) and BMV \cite{CarrascoPNAS06,MichelPNAS06,SnijderNC13,VaughanJV14,NiJMB12}. Although these experiments reveal an amazing diversity of mechanical properties of biological particles, experimental results are difficult to interpret without theoretical modeling. What types of mechanical excitations drive the particle deformation and collapse? What determines the mechanical limit of the particle --- the critical forces and critical deformations? Why is the initial portion of the $FX$ spectra weakly non-linear? Why do the $FX$ spectra for the same particle differ from one measurement to another? This points to the stochastic nature of collapse transitions, but what defines the likelihood of structural collapse at a given force load? \\
\indent A number of theoretical approaches have been designed to describe the dynamics of virus particles, including: finite element analysis \cite{GibbonsBJ13,KlugPRL12}, normal mode analysis \cite{TamaJMB05}, elastic network modeling \cite{YangBJ09}, atomistic MD and coarse-grained simulations \cite{ZinkBJ10,ArkhipovBJ09,CieplakJCP10}, and other approaches \cite{MayBJ11}. Here we take a step further to develop an analytically tractable model for meaningful interpretation of the force-deformation spectral lineshapes available from single-particle nanomanipulation experiments. The theory links the slope, critical force, and the critical deformation of the $FX$ curve with the physical characteristics of the structure, geometry and overall shape of the particle and indenter. We identify the types of mechanical excitations (degrees of freedom), which contribute to the particle deformation (indentation depth) $X$, by analyzing the structure and potential energy changes in the CCMV particle using our results of nanoindentations \textit{in silico}. The methodology of nanoindentation \textit{in silico} is described in Ref.\cite{KononovaBJ13}; see Supplementary Material (SM; Figs. S1-S3). We formulate and apply the Fluctuating Nonlinear Spring (FNS) model to characterize the $FX$ spectra for the CCMV, AdV and TrV particles obtained as described in Refs.\cite{SnijderJV13, SnijderNC13, SnijderMicron12}. \\
\indent \textit{Degrees of freedom} --- In dynamic force-ramp assays $f(t)$$=$$ \kappa \nu_ft$, an indenter (cantilever tip) compresses a particle (Fig.~\ref{fig:fig1}, S3) creating a physical contact between them. The force loads the particle mechanically over time $t$ with the force-loading rate $\kappa\nu_f$ ($\kappa$ and $\nu_f$ are the cantilever spring constant and velocity). For small $f$, the mechanical energy is localized to the virion surface under the tip, which results in the tip and particle undergoing normal displacements $u_{tip}$ and $u_{par}$ corresponding to the Hertzian deformation $x_H$$=$$u_{tip}$$+$$u_{par}$. Typically, $u_{tip}$$\ll$$u_{par}$, therefore $x_H$$=$$u_{par}$. The mechanical energy gradually fills the particle structure stressing the side portions of the structure undergoing bending deformations $x_b$ (Fig.~\ref{fig:fig1}). The total deformation is the sum of Hertzian and bending deformations, i.e. $X$$=$$x_H$$+$$x_b$, and the deformation force $F$ of the particle (restoring force exerted on the tip by the particle) is a bivariate function, $F(X$$=$$x_H$$+$$x_b)$$=$$F_H(x_H)$$+$$F_b(x_b)$. We quantified the amplitude of Hertzian deformation $x_H$$\approx$$3$ nm and bending deformations $x_b$$\approx$$4.3$ nm using the simulation output for the CCMV (Fig.~\ref{fig:fig1}). \\
\indent The Hertz model \cite{Landau,Johnson} accounts for the force due to local curvature change $F_H(x_H)$ (Fig.~\ref{fig:fig1}(a)),
\begin{equation}\label{eq:fh}
F_H(x_H)=1/D_H (R_{par}R_{tip}/(R_{par}+R_{tip}))^{1/2} \cdot x_H^{3/2}
\end{equation}
where $R_{par}$ and $R_{tip}$ are the radii of the particle and the tip and $D_H$$=$$0.75((1-\sigma_H^2)/E_H + (1-\sigma_{tip}^2)/E_{tip})$. $E_H$ and $E_{tip}$ are the Young's moduli and $\sigma_H$ and $\sigma_{tip}$ are the Poisson's ratios for the particle and the tip deformations, respectively. Since $E_{tip}$$\gg$$E_H$, $D_H$$=$$0.75(1-\sigma_H^2)/E_H$. To describe the bending deformations $F_b(x_b)$, we divide the side portion of the particle structure into vertical curved beams of length $L$ (Fig.~\ref{fig:fig1}(d)). For a spherical particle with thickness $r$, the total number of beams is $N$$=$$2\pi R_{par}/r$. For small $x_b$ (Fig.~\ref{fig:fig1}), the potential energy change is $E_bI/2\int_L(\kappa(x_b,l)$$-$$\kappa_0)^2dl$ \cite{Landau,Timoshenko}, where $\kappa_0$ and $\kappa(x_b,l)$ are the initial and current curvature of the beam element $dl$ ($0$$\leq$$l$$\leq$$R_{par}$$-$$x_b/2$) and $E_bI$ is the flexural rigidity. With the beam shape function $q(x_b,l)$$=$$(R_{par}$$+$$x_b/2)(1-l^2/(R_{par}-x_b/2)^2)^{1/2}$ the curvature of the beam is given by $\kappa(x_b,l)$$=$$q^{\prime\prime}(x_b,l)/(1$$+$$(q^\prime(x_b,l))^2)^{3/2}$ ($q^{\prime}$ and $q^{\prime\prime}$ are the first and second derivatives of $q$ with respect to $l$). By performing the integration we obtain the expression for the beam bending energy, which upon differentiation with respect to $x_b$, gives the beam-bending force. Expanding the resulting expression in Taylor series in powers of $x_b$ and retaining the linear term we obtain:
\begin{equation}\label{eq:fbexp}
f_b(x_b) \cong 9 E_b I \pi / 8 R_{par}^3 \cdot x_b
\end{equation}
Combining the contributions from all $N$ beams and adding together Eqs.(\ref{eq:fh}) and (\ref{eq:fbexp}), we obtain the deformation force $\tilde F(x_H,x_b)$$=$$F_H(x_H)$$+$$F_b(x_b)$$=$$k_H x_H^{3/2}$$+$$Nk_b x_b$, where $k_H$ $=$ $(R_{par}R_{tip}/(R_{par}+R_{tip})^{1/2})/D_H$ (Eq.(\ref{eq:fh})) and $k_b$$=$$9E_b I \pi /(8 R_{par}^3)$ is the beam spring constant. In agreement with the results of \textit{in silico} indentation of CCMV (Fig.~\ref{fig:fig2}), $F(x_H, x_b)$ predicts that the initial portion of $FX$ curves, where Hertzian deformation dominates ($X$$\approx$$x_H$), is weakly nonlinear. The Hertzian nonlinearity expressed in $\tilde F(x_H,x_b)$ was indeed observed in recent experiments on thick-shelled particles \cite{MichelPNAS06}. However, $\tilde F(x_H,x_b)$ does not capture the catastrophic force drop (Fig.~\ref{fig:fig2}) because the theory lacks a description of structural damage. \\
\indent \textit{Fluctuating Nonlinear Spring (FNS) model} --- We represent a particle by a collection of $N$ identical beams interacting with an indenter through a Hertzian cushion. Each $i$-th beam undergoes the equal elastic deformation $x_{bi}$$=$$x_b$ with the spring constant ($i$$=$$1, 2, \dots , N$) until it fails mechanically when the load on the beam reaches some critical value $f_{bi}^*$. The spherical geometry of a virus particle dictates the parallel arrangement of beams with the spring $K_b$$=$$\sum_{i=1}^{N}k_{bi}$$=$$Nk_b$. There are $n$ (or $N$$-$$n$) beams that have failed (or survived), and the actual bending force is $F_b(x_b)$$=$$k_b(N-n)x_b$$=$$K_b x_b (1-n/N)$. We define the random variables associated with the probability of damage $d$$=$$n/N$ and survival $s$$=$$(N - n)/N$$=$$1 - d$ of the collection of beams ($0$$\leq$$d$, $s$$\leq$$1$). In the continuous limit, $d$ and $s$ are described by the pdfs $p(F_b)$, $d(F_b)$$=$$Prob(F_b$$=$$K_bx_b)$$=$$\int_{0}^{F_b}p(F_b^\prime)dF_b^\prime$ and $s(F_b)$$=$$1 - d(F_b)$$=$$\int_{F_b}^{\infty}p(F_b^\prime)dF_b^\prime$. With the particle damage accounted for, the deformation force becomes:
\begin{equation}\label{eq:damage}
F(x_H,x_b) = k_H x_H^{3/2} + K_b x_b s(x_b)
\end{equation}
Here, $s$ can be estimated from computer simulations using the structural similarity between a given structure and the initial state $\chi(x_b)$$=$$(2N(N-1))^{-1}\sum{\Theta(|r_{ij}(x_b)-r_{ij}(0)|-\beta r_{ij}(0))}$. In the Heaviside step function, $r_{ij}(x_b)$ and $r_{ij}(0)$ are the distances between the $i$-th and $j$-th amino acids in the given and initial structure, respectively, and $\beta$$=$$0.2$ is the tolerance for the distance change. Because $\chi(x_b)$ ranges from $\chi$$=$$1$ (fully similar structures) to $\chi$$=$$0$ (completely dissimilar structures) and changes in $s(x_b)$ are commensurate with structure changes, we have $s(x_b)$$\approx$$\chi(x_b)$. The $\chi$-based estimate of $s(x_b)$ from force-deformation \textit{in silico} of CCMV shows that this metric can be used to inform the modeling of $p(F_b)$ (Fig.~\ref{fig:fig2}).\\
\indent The transition to the collapsed state occurs when all beams have failed and so, the longest lasting beam determines the collapse onset at the critical force $F^{col}$ (critical deformation $X^{col}$). Hence, the statistics of the extreme (maximum) force determines the beams' failure. We used the Weibull distribution \cite{Gumbel}
\begin{eqnarray}\label{eq:weibull}
d(x_b) = 1 - e^{-(K_b x_b/F_b^*)^m} \text{, } s(x_b) = e^{-(K_b x_b/F_b^*)^m}
\end{eqnarray}
with the shape parameter $m$ (failure rate), and scale parameter $F_b^*$ which can be understood by using the condition of maximum force. By substituting $s(x_b)$ into $F_b(x_b)$ in Eq.(\ref{eq:damage}), by differentiating this expression with respect to $x_b$, and by setting $dF_b/dx_b$$=$$0$, we obtain $F_b^*$$=$$K_b x_b^*\sqrt[m]{m}$ ($x_b^*$ is the critical beam deformation). By substituting $F_b^*$ into the expression for $F_b(x_b)$ and into Eq.(\ref{eq:weibull}), we obtain the beam-bending force threshold $F_b^{col}$$=$$F_b^* / \sqrt[m]{em}$$=$$K_b x_b^*/\sqrt[m]{e}$ and the critical value of collapse probability $d^{col}$$=$$1 - s^{col}$$=$$1 - 1/\sqrt[m]{e}$. \\
\indent By substituting Eq.(\ref{eq:weibull}) for $s(x_b)$ in Eq.(\ref{eq:damage}), we obtain the main result of this Letter:
\begin{equation}\label{eq:final}
F(x_H,x_b) = k_H x_H^{3/2} + K_b x_b e^{-(K_bx_b/F_b^*)^m}
\end{equation}
Eq.(\ref{eq:final}) shows that a biological particle behaves as a nonlinear spring. The beams' bending starts as elastic $(Nk_b$), but becomes stochastic near the collapse transition, where the particle mechanical resistance fluctuates, thus explaining the variability of $F^{col}$ and $X^{col}$ in the $FX$ spectra (Figs.~\ref{fig:fig2}, \ref{fig:fig3}). Hence, the uniaxial forced deformation of a biological particle can be represented by the mechanical evolution of a fluctuating nonlinear spring (FNS). \\
\indent \textit{Application of FNS model} --- In the experiment, $F$ is measured as a function of the sum $X$$=$$x_H+x_b$. A particular realization of the forced deformation process ($FX$ trajectory) is a stochastic path on the 2D surface $F(x_H,x_b)$ (Fig.~\ref{fig:fig3}). For slow loading, when the particle structure equilibrates on a timescale faster than the rate of force change, the most dominant path is the equilibrium path $F_{eq}(X_{eq})$. Using slow cantilever velocities ($\nu_f$$=$$0.06$$-$$1.0$ $\mu$m/s) allows us to use this quasi-equilibrium argument. The equilibrium force $F_{eq}$ can then be determined from the requirement that the deformation force (deformation energy) attains the minimum. Finding $F_{eq}$ is equivalent to finding a pair $x_H$ and $x_b$ that minimizes $F(x_H,x_b)$ subject to the constraint, $X$$=$$x_H+x_b$ (Fig.~\ref{fig:fig3}), which can be solved using Lagrange multipliers (SM). \\
\indent The average simulated spectra for CCMV are compared with the theoretical curves in Fig.~\ref{fig:fig2}
(simulated force-deformation spectra for the CCMV particle are in Fig. S4(a)). To find the best fit, we employed two methods. The exact method is based on Eq.(\ref{eq:final}) and uses Lagrange multipliers to find $x_H$ and $x_b$. This requires solving for $x_b$ the non-linear equation $a_1x_b^{4m}+a_2x_b^{3m}+a_3x_b^{2m}+a_4x_b^{m}+a_5x_b+a_6=0$, where $a_1$$=$$mK_b^2(K_b/F_b^*)^{4m}$, $a_2$$=$$-2m(1+m)K_b^2(K_b/F_b^*)^{3m}$, $a_3$$=$$(1+4m+m^2)K_b^2(K_b/F_b^*)^{2m}$, $a_4$$=$$-2(1+m)K_b^2(K_b/F_b^*)^m$, $a_5$$=$$9/4k_H^2$, and $a_6$$=$$K_b^2-9/4k_H^2X$. Then, $x_H$ is obtained as $x_H$$=$$X$$-$$x_b$. In the piece-wise approximate method (Fig.~\ref{fig:fig2}), the spectrum is divided into the Hertzian-deformation-dominated initial regime I: $X$$\approx$$x_H$ ($x_b$$\approx$$0$) and $F$$\approx$$F_H$$=$$k_HX^{3/2}$; and the transition regime II (non-monotonic part of $FX$ curve): $X$$\approx$$x_b$ and $F$$\approx$$F_b$$=$$K_bXs(X)$. We calculate $F_H$ in regime I for $X$$\approx$$x_H$$<$$x_H^{max}$; $x_H^{max}$ is obtained using Lagrange multipliers and setting $s(x_b)$$=$$1$. In regime II, we use $F(x_H^{max},x_b)$$=$$k_H(x_H^{max})^{3/2}+K_b(x_b-x_H^{max}\exp\left[-(K_b(x_b-x_H^{max})/F_b^*)^m\right])$ for $X$$\approx$$x_b$$>$$x_H^{max}$. \\
\indent The agreement between the simulated force-deformation spectra and theoretical $FX$ curves for the CCMV particle is almost quantitative (Fig.~\ref{fig:fig2}); model parameters obtained using both methods are very close (Table I). For all symmetry types, the Hertzian excitation is softer than the beam-bending ($k_H$$<$$K_b$) implying smaller Young's modulus for Hertzian deformation, $E_H$$<$$E_b$, which is why the Hertzian degree of freedom is excited first (Fig.~\ref{fig:fig1}). After the Hertzian force reached the maximum $F_H^{max}$$=$$k_H(x_H^{max})^{3/2}$ at $X$$\approx$$x_H$$=$$x_H^{max}$, a subsequent force increase excites the beam-bending degrees of freedom and $x_H$ ($x_b$) decreases (increases). Hence, physical properties of the particle are dynamic as the nature of its mechanical response changes with increasing $X$ from Hertzian to beam-bending deformation, because the actual stiffness of beams is degraded with increasing $x_b$ as $K_b\exp\left[-(K_bx_b/F_b^*)^m\right]$. \\
\indent The FNS model accounts for the dependence of mechanical response of biological particles on particle and indenter geometries \cite{KononovaBJ13}. Since the information about the particle/indenter size is contained in $k_H$, geometric effects are important in the Hertzian deformation regime. When $m$$>$$1$ ($m$$<$$1$), the beam failure events become more frequent (less frequent) with increasing $x_b$. For CCMV, $1.8$$<$$m$$<$$2.1$ (Table I), which indicates that the beam-failure rate increases with $X$. When first $n$ beams fail, the compressive force is redistributed among the remaining $N - n$ beams, and each surviving beam experiences an increasingly larger tension with every next beam failure, which accelerates the failure frequency. The beams fail not when $f$$>$$F_b^*$, but under smaller forces $F_b^{col}$$=$$F_b^*/\sqrt[m]{em}$ (for $m$$\approx$$2$, $F_b^{col}$$\approx$$0.43F_b^*$). \\
\indent The AFM-based measurements for the empty CCMV shell, empty TrV capsid, full TrV virion and full AdV virion are in Figs. S4, S5. Theoretical fit to the experimental average $FX$ curves for these particles
shows that their deformations are well described by the FNS model (Fig.~\ref{fig:fig3}). The obtained Young's moduli for Hertzian deformation are uniformly smaller (~$10$$-$$100$ MPa) than the Young's moduli for bending deformation (Giga-Pascal range; Table I). This correlates with the $FX$ curves being only weakly nonlinear (Fig.~\ref{fig:fig3}). There are small variations in the model parameters for the AdV virion due to force application along different symmetry axes. This correlates with our similar findings for the CCMV shell, and implies that the symmetry of local arrangements of capsomer repeats at the point of indentation influences its mechanics \cite{KononovaBJ13}. The failure rates are found to be accelerating with deformation ($m$$>$$1$). This fully reflects the conditions of application of compressive force, which increases linearly in time speeding up beams' failure events. Parameters for empty and ssRNA-loaded TrV capsids indicate that the difference in particle stiffness is largely due to an increase in the Young's modulus for Hertzian deformation $E_H$$=$$0.03$ GPa (empty TrV) vs $0.14$ GPa (full TrV), which suggests that local indentations are resisted in ssRNA-filled particles. These results fit with the previously observed uniaxial deformation of RNA-filled TrV into an oblate sphere to maximize the volume available to pack the genome \cite{SnijderNC13}. Hence, confining the large ssRNA genome inside the small particle volume builds internal pressure resisting local indentation. \\
\indent Previously, the 3D Young’s modulus of the capsid material was estimated using a thin shell theory \cite{RoosNatPhys10,MichelPNAS06,SnijderNC13,Landau}. This assumption is valid for some bacteriophage capsids, but in CCMV and TrV the shell thickness is comparable with the virion radius. The FNS model accounts for compression of the protein layer under the tip. In the FNS model, the beam-bending modulus ($E_b$) is roughly equivalent to the 3D Young's modulus in the thin shell theory. It is estimated at $\sim$$0.85$ GPa (experiment) and $\sim$$0.4$$-$$0.5$ GPa (simulations) for the empty CCMV capsid (Table I). These are similar to yet larger than the values of $0.15$$-$$0.30$ GPa obtained with the thin shell theory \cite{RoosNatPhys10, MichelPNAS06} and $0.28$$-$$0.36$ GPa from finite-element analysis ($\sim$$0.25$ GPa) \cite{GibbonsPRE07}, but they disagree with the estimates from several computer modeling studies ($0.08$$-$$0.09$ GPa) \cite{CieplakJCP10,MayBJ11}. In the  study based on spherical harmonics \cite{MayBJ11}, multiple deformation modes have also been observed, corresponding to equilibrium deformations of the polar regions (tip/surface contact area in FNS model) and the side wall (beams in FNS model) of the shell. For the empty TrV capsid, we obtain $E_b$$\approx$$0.9$ GPa (Table I) whereas the thin shell theory gives a smaller value of $\sim$$0.5$ GPa. The lower values of the 3D Young's modulus result from attributing the softer Hertzian deformation mode to bending of the capsid shell in the thin shell theory. Indeed, for CCMV and TrV shells, the thin shell theory estimates $0.15$$-$$0.30$ GPa and $0.5$ GPa are between the values of $E_H$$=$$0.02$$-$$0.03$ GPa and $E_b$$=$$0.85$$-$$0.95$ GPa from FNS modeling (Table I). \\
\indent We used parameters of the FNS model and positions of the force maximum at $X^{col}$$=$$x_H^*+x_b^*$ from the $FX$ spectra (Figs.~\ref{fig:fig2}, \ref{fig:fig3}) to predict the critical force for collapse, $F^{col}$$=$$F_H(x_H^*)+F_b^{col}(x_b^*)$$=$$k_Hx_H^{*3/2}+K_bx_b^*/\sqrt[m]{e}$. The obtained values of $F^{col}$ (Table I) agree with their counterparts extracted from the $FX$ curves (Figs.~\ref{fig:fig2}, \ref{fig:fig3}), which validates the model. The analytically tractable FNS model uniquely combines the elements of continuum mechanics and statistics of extremes to accurately describe the mechanical deformation and collapse of biological particles. The model can also be extended to characterize the biological particles with other regular geometries, e.g. microtubule polymers (cylinder) \cite{DePabloPRL03, KononovaJACS14}.
%
\begin{acknowledgments}
This work was supported by the American Heart Association (grant-in-aid  15GRNT23150000 to VB) and by VIDI grant of the Nederlandse Organisatie voor Wetenschappelijk Onderzoek (to WHR).
\end{acknowledgments}
%

\begin{thebibliography}{10}%
\makeatletter
\providecommand \@ifxundefined [1]{%
 \ifx #1\undefined \expandafter \@firstoftwo
 \else \expandafter \@secondoftwo
\fi
}%
\providecommand \@ifnum [1]{%
 \ifnum #1\expandafter \@firstoftwo
 \else \expandafter \@secondoftwo
\fi
}%
\providecommand \enquote [1]{``#1''}%
\providecommand \bibnamefont  [1]{#1}%
\providecommand \bibfnamefont [1]{#1}%
\providecommand \citenamefont [1]{#1}%
\providecommand\href[0]{\@sanitize\@href}%
\providecommand\@href[1]{\endgroup\@@startlink{#1}\endgroup\@@href}%
\providecommand\@@href[1]{#1\@@endlink}%
\providecommand \@sanitize [0]{\begingroup\catcode`\&12\catcode`\#12\relax}%
\@ifxundefined \pdfoutput {\@firstoftwo}{%
 \@ifnum{\z@=\pdfoutput}{\@firstoftwo}{\@secondoftwo}%
}{%
 \providecommand\@@startlink[1]{\leavevmode}%
 \providecommand\@@endlink[0]{}%
}{%
 \providecommand\@@startlink[1]{%
  \leavevmode
  \pdfstartlink
   attr{/Border[0 0 1 ]/H/I/C[0 1 1]}%
   user{/Subtype/Link/A<</Type/Action/S/URI/URI(#1)>>}%
  \relax
 }%
 \providecommand\@@endlink[0]{\pdfendlink}%
}%
\providecommand \url  [0]{\begingroup\@sanitize \@url }%
\providecommand \@url [1]{\endgroup\@href {#1}{\urlprefix}}%
\providecommand \urlprefix [0]{URL }%
\providecommand \Eprint[0]{\href }%
\@ifxundefined \urlstyle {%
  \providecommand \doi [1]{doi:\discretionary{}{}{}#1}%
}{%
  \providecommand \doi [0]{doi:\discretionary{}{}{}\begingroup
  \urlstyle{rm}\Url }%
}%
\providecommand \doibase [0]{http://dx.doi.org/}%
\providecommand \Doi[1]{\href{\doibase#1}}%
\providecommand \bibAnnote [3]{%
  \BibitemShut{#1}%
  \begin{quotation}\noindent
    \textsc{Key:}\ #2\\\textsc{Annotation:}\ #3%
  \end{quotation}%
}%
\providecommand \bibAnnoteFile [2]{%
  \IfFileExists{#2}{\bibAnnote {#1} {#2} {\input{#2}}}{}%
}%
\providecommand \typeout [0]{\immediate \write \m@ne }%
\providecommand \selectlanguage [0]{\@gobble}%
\providecommand \bibinfo [0]{\@secondoftwo}%
\providecommand \bibfield [0]{\@secondoftwo}%
\providecommand \translation [1]{[#1]}%
\providecommand \BibitemOpen[0]{}%
\providecommand \bibitemStop [0]{}%
\providecommand \bibitemNoStop [0]{.\EOS\space}%
\providecommand \EOS [0]{\spacefactor3000\relax}%
\providecommand \BibitemShut [1]{\csname bibitem#1\endcsname}%
\bibitem{RoosNatPhys10}%
  \BibitemOpen
  \bibfield{author}{%
  \bibinfo {author} {\bibfnamefont{W.~H.}\ \bibnamefont{Roos}}, \bibinfo
  {author} {\bibfnamefont{R.}~\bibnamefont{Bruinsma}},\ and\ \bibinfo {author}
  {\bibfnamefont{G.~J.~L.}\ \bibnamefont{Wuite}},\ }%
  \bibfield{journal}{%
  \bibinfo {journal} {Nat.\ Phys.}\ }%
  \textbf{\bibinfo {volume} {6}},\ \bibinfo {pages} {733} (\bibinfo {year}
  {2010})%
  \bibAnnoteFile{NoStop}{RoosNatPhys10}%
\bibitem{IvanovskaPNAS04}%
  \BibitemOpen
  \bibfield{author}{%
  \bibinfo {author} {\bibfnamefont{I.~L.}\ \bibnamefont{Ivanovska}}, \bibinfo
  {author} {\bibfnamefont{P.~J.}\ \bibnamefont{de~Pablo}}, \bibinfo {author}
  {\bibfnamefont{B.}~\bibnamefont{Ibarra}}, \bibinfo {author}
  {\bibfnamefont{G.}~\bibnamefont{Sgalari}}, \bibinfo {author}
  {\bibfnamefont{F.~C.}\ \bibnamefont{MacKintosh}}, \bibinfo {author}
  {\bibfnamefont{J.~L.}\ \bibnamefont{Carrascosa}}, \bibinfo {author}
  {\bibfnamefont{C.~F.}\ \bibnamefont{Schmidt}},\ and\ \bibinfo {author}
  {\bibfnamefont{G.~J.~L.}\ \bibnamefont{Wuite}},\ }%
  \bibfield{journal}{%
  \bibinfo {journal} {Proc.\ Natl.\ Acad.\ Sci.\ USA}\ }%
  \textbf{\bibinfo {volume} {101}},\ \bibinfo {pages} {7600} (\bibinfo {year}
  {2004})%
  \bibAnnoteFile{NoStop}{IvanovskaPNAS04}%
\bibitem{HernandoPerezNano14}%
  \BibitemOpen
  \bibfield{author}{%
  \bibinfo {author} {\bibfnamefont{M.}~\bibnamefont{Hernando-P$\grave{e}$rez}},
  \bibinfo {author} {\bibfnamefont{E.}~\bibnamefont{Pascual}}, \bibinfo
  {author} {\bibfnamefont{M.}~\bibnamefont{Aznar}}, \bibinfo {author}
  {\bibfnamefont{A.}~\bibnamefont{Ionel}}, \bibinfo {author}
  {\bibfnamefont{J.~R.}\ \bibnamefont{Castón}}, \bibinfo {author}
  {\bibfnamefont{A.}~\bibnamefont{Luque}}, \bibinfo {author}
  {\bibfnamefont{J.~L.}\ \bibnamefont{Carrascosa}}, \bibinfo {author}
  {\bibfnamefont{D.}~\bibnamefont{Reguera}},\ and\ \bibinfo {author}
  {\bibfnamefont{P.~J.}\ \bibnamefont{de~Pablo}},\ }%
  \bibfield{journal}{%
  \bibinfo {journal} {Nanoscale}\ }%
  \textbf{\bibinfo {volume} {6}},\ \bibinfo {pages} {2702} (\bibinfo {year}
  {2014})%
  \bibAnnoteFile{NoStop}{HernandoPerezNano14}%
\bibitem{RoosPNAS12}%
  \BibitemOpen
  \bibfield{author}{%
  \bibinfo {author} {\bibfnamefont{W.~H.}\ \bibnamefont{Roos}}, \bibinfo
  {author} {\bibfnamefont{I.}~\bibnamefont{Gertsman}}, \bibinfo {author}
  {\bibfnamefont{E.~R.}\ \bibnamefont{May}}, \bibinfo {author}
  {\bibfnamefont{C.~L.}\ \bibnamefont{{Brooks 3rd}}}, \bibinfo {author}
  {\bibfnamefont{J.~E.}\ \bibnamefont{Johnson}},\ and\ \bibinfo {author}
  {\bibfnamefont{G.~J.~L.}\ \bibnamefont{Wuite}},\ }%
  \bibfield{journal}{%
  \bibinfo {journal} {Proc.\ Natl.\ Acad.\ Sci.\ USA}\ }%
  \textbf{\bibinfo {volume} {109}},\ \bibinfo {pages} {2342} (\bibinfo {year}
  {2012})%
  \bibAnnoteFile{NoStop}{RoosPNAS12}%
\bibitem{KolBJ07}%
  \BibitemOpen
  \bibfield{author}{%
  \bibinfo {author} {\bibfnamefont{M.}~\bibnamefont{Kol}}, \bibinfo {author}
  {\bibfnamefont{Y.}~\bibnamefont{Shi}}, \bibinfo {author}
  {\bibfnamefont{M.}~\bibnamefont{Tsvitov}}, \bibinfo {author}
  {\bibfnamefont{D.}~\bibnamefont{Barlam}}, \bibinfo {author}
  {\bibfnamefont{R.~Z.}\ \bibnamefont{Shneck}}, \bibinfo {author}
  {\bibfnamefont{M.~S.}\ \bibnamefont{Kay}},\ and\ \bibinfo {author}
  {\bibfnamefont{I.}~\bibnamefont{Rousso}},\ }%
  \bibfield{journal}{%
  \bibinfo {journal} {Biophys.\ J.}\ }%
  \textbf{\bibinfo {volume} {92}},\ \bibinfo {pages} {1777} (\bibinfo {year}
  {2007})%
  \bibAnnoteFile{NoStop}{KolBJ07}%
\bibitem{PerezBernaJBC12}%
  \BibitemOpen
  \bibfield{author}{%
  \bibinfo {author} {\bibfnamefont{A.~J.}\
  \bibnamefont{P$\acute{e}$rez-Bern$\acute{a}$}}, \bibinfo {author}
  {\bibfnamefont{A.}~\bibnamefont{Ortega-Esteban}}, \bibinfo {author}
  {\bibfnamefont{R.}~\bibnamefont{Menendez-Conejero}}, \bibinfo {author}
  {\bibfnamefont{D.~C.}\ \bibnamefont{Winkler}}, \bibinfo {author}
  {\bibfnamefont{M.}~\bibnamefont{Menendez}}, \bibinfo {author}
  {\bibfnamefont{A.~C.}\ \bibnamefont{Steven}}, \bibinfo {author}
  {\bibfnamefont{S.~J.}\ \bibnamefont{Flint}}, \bibinfo {author}
  {\bibfnamefont{P.~J.}\ \bibnamefont{de~Pablo}},\ and\ \bibinfo {author}
  {\bibfnamefont{C.~S.}\ \bibnamefont{Martin}},\ }%
  \bibfield{journal}{%
  \bibinfo {journal} {J.\ Biol.\ Chem.}\ }%
  \textbf{\bibinfo {volume} {287}},\ \bibinfo {pages} {31582} (\bibinfo {year}
  {2012})%
  \bibAnnoteFile{NoStop}{PerezBernaJBC12}%
\bibitem{RoosPNAS09}%
  \BibitemOpen
  \bibfield{author}{%
  \bibinfo {author} {\bibfnamefont{W.~H.}\ \bibnamefont{Roos}}, \bibinfo
  {author} {\bibfnamefont{K.}~\bibnamefont{Radtke}}, \bibinfo {author}
  {\bibfnamefont{E.}~\bibnamefont{Kniesmeijer}}, \bibinfo {author}
  {\bibfnamefont{H.}~\bibnamefont{Geertsema}}, \bibinfo {author}
  {\bibfnamefont{B.}~\bibnamefont{Sodeik}},\ and\ \bibinfo {author}
  {\bibfnamefont{G.~J.~L.}\ \bibnamefont{Wuite}},\ }%
  \bibfield{journal}{%
  \bibinfo {journal} {Proc.\ Natl.\ Acad.\ Sci.\ USA}\ }%
  \textbf{\bibinfo {volume} {106}},\ \bibinfo {pages} {9673} (\bibinfo {year}
  {2009})%
  \bibAnnoteFile{NoStop}{RoosPNAS09}%
\bibitem{SnijderJV13}%
  \BibitemOpen
  \bibfield{author}{%
  \bibinfo {author} {\bibfnamefont{J.}~\bibnamefont{Snijder}}, \bibinfo
  {author} {\bibfnamefont{V.~S.}\ \bibnamefont{Reddy}}, \bibinfo {author}
  {\bibfnamefont{E.~R.}\ \bibnamefont{May}}, \bibinfo {author}
  {\bibfnamefont{W.~H.}\ \bibnamefont{Roos}}, \bibinfo {author}
  {\bibfnamefont{G.~R.}\ \bibnamefont{Nemerow}},\ and\ \bibinfo {author}
  {\bibfnamefont{G.~J.~L.}\ \bibnamefont{Wuite}},\ }%
  \bibfield{journal}{%
  \bibinfo {journal} {J.\ Virol.}\ }%
  \textbf{\bibinfo {volume} {87}},\ \bibinfo {pages} {2756} (\bibinfo {year}
  {2013})%
  \bibAnnoteFile{NoStop}{SnijderJV13}%
\bibitem{CarrascoPNAS06}%
  \BibitemOpen
  \bibfield{author}{%
  \bibinfo {author} {\bibfnamefont{C.}~\bibnamefont{Carrasco}}, \bibinfo
  {author} {\bibfnamefont{A.}~\bibnamefont{Carreira}}, \bibinfo {author}
  {\bibfnamefont{I.~A.~T.}\ \bibnamefont{Schaap}}, \bibinfo {author}
  {\bibfnamefont{P.~A.}\ \bibnamefont{Serena}}, \bibinfo {author}
  {\bibfnamefont{J.}~\bibnamefont{G$\acute{o}$mez-Herrero}}, \bibinfo {author}
  {\bibfnamefont{M.~G.}\ \bibnamefont{Mateu}},\ and\ \bibinfo {author}
  {\bibfnamefont{P.~J.}\ \bibnamefont{de~Pablo}},\ }%
  \bibfield{journal}{%
  \bibinfo {journal} {Proc.\ Natl.\ Acad.\ Sci.\ USA}\ }%
  \textbf{\bibinfo {volume} {103}},\ \bibinfo {pages} {13706} (\bibinfo {year}
  {2006})%
  \bibAnnoteFile{NoStop}{CarrascoPNAS06}%
\bibitem{MichelPNAS06}%
  \BibitemOpen
  \bibfield{author}{%
  \bibinfo {author} {\bibfnamefont{J.~P.}\ \bibnamefont{Michel}}, \bibinfo
  {author} {\bibfnamefont{I.~L.}\ \bibnamefont{Ivanovska}}, \bibinfo {author}
  {\bibfnamefont{M.~M.}\ \bibnamefont{Gibbons}}, \bibinfo {author}
  {\bibfnamefont{W.~S.}\ \bibnamefont{Klug}}, \bibinfo {author}
  {\bibfnamefont{C.~M.}\ \bibnamefont{Knobler}}, \bibinfo {author}
  {\bibfnamefont{G.~J.~L.}\ \bibnamefont{Wuite}},\ and\ \bibinfo {author}
  {\bibfnamefont{C.~F.}\ \bibnamefont{Schmidt}},\ }%
  \bibfield{journal}{%
  \bibinfo {journal} {Proc.\ Natl.\ Acad.\ Sci.\ USA}\ }%
  \textbf{\bibinfo {volume} {103}},\ \bibinfo {pages} {6184} (\bibinfo {year}
  {2006})%
  \bibAnnoteFile{NoStop}{MichelPNAS06}%
\bibitem{SnijderNC13}%
  \BibitemOpen
  \bibfield{author}{%
  \bibinfo {author} {\bibfnamefont{J.}~\bibnamefont{Snijder}}, \bibinfo
  {author} {\bibfnamefont{C.}~\bibnamefont{Uetrecht}}, \bibinfo {author}
  {\bibfnamefont{R.}~\bibnamefont{Rose}}, \bibinfo {author}
  {\bibfnamefont{R.}~\bibnamefont{Sanchez}}, \bibinfo {author}
  {\bibfnamefont{G.}~\bibnamefont{Marti}}, \bibinfo {author}
  {\bibfnamefont{J.}~\bibnamefont{Agirre}}, \bibinfo {author}
  {\bibfnamefont{D.~M.}\ \bibnamefont{Gu$\acute{e}$rin}}, \bibinfo {author}
  {\bibfnamefont{G.~J.~L.}\ \bibnamefont{Wuite}}, \bibinfo {author}
  {\bibfnamefont{A.~J.~R.}\ \bibnamefont{Heck}},\ and\ \bibinfo {author}
  {\bibfnamefont{W.~H.}\ \bibnamefont{Roos}},\ }%
  \bibfield{journal}{%
  \bibinfo {journal} {Nature\ Chemistry}\ }%
  \textbf{\bibinfo {volume} {5}},\ \bibinfo {pages} {502} (\bibinfo {year}
  {2013})%
  \bibAnnoteFile{NoStop}{SnijderNC13}%
\bibitem{VaughanJV14}%
  \BibitemOpen
  \bibfield{author}{%
  \bibinfo {author} {\bibfnamefont{R.}~\bibnamefont{Vaughan}}, \bibinfo
  {author} {\bibfnamefont{B.}~\bibnamefont{Tragesser}}, \bibinfo {author}
  {\bibfnamefont{P.}~\bibnamefont{Ni}}, \bibinfo {author}
  {\bibfnamefont{X.}~\bibnamefont{Ma}}, \bibinfo {author}
  {\bibfnamefont{B.}~\bibnamefont{Dragnea}},\ and\ \bibinfo {author}
  {\bibfnamefont{C.}~\bibnamefont{Kao}},\ }%
  \bibfield{journal}{%
  \bibinfo {journal} {J.\ Virol.}\ }%
  \textbf{\bibinfo {volume} {88}},\ \bibinfo {pages} {6483} (\bibinfo {year}
  {2014})%
  \bibAnnoteFile{NoStop}{VaughanJV14}%
\bibitem{NiJMB12}%
  \BibitemOpen
  \bibfield{author}{%
  \bibinfo {author} {\bibfnamefont{P.}~\bibnamefont{Ni}}, \bibinfo {author}
  {\bibfnamefont{Z.}~\bibnamefont{Wang}}, \bibinfo {author}
  {\bibfnamefont{X.}~\bibnamefont{Ma}}, \bibinfo {author}
  {\bibfnamefont{N.~C.}\ \bibnamefont{Das}}, \bibinfo {author}
  {\bibfnamefont{P.}~\bibnamefont{Sokol}}, \bibinfo {author}
  {\bibfnamefont{W.}~\bibnamefont{Chiu}}, \bibinfo {author}
  {\bibfnamefont{B.}~\bibnamefont{Dragnea}}, \bibinfo {author}
  {\bibfnamefont{M.}~\bibnamefont{Hagan}},\ and\ \bibinfo {author}
  {\bibfnamefont{C.}~\bibnamefont{Kao}},\ }%
  \bibfield{journal}{%
  \bibinfo {journal} {J.\ Mol.\ Biol.}\ }%
  \textbf{\bibinfo {volume} {419}},\ \bibinfo {pages} {284} (\bibinfo {year}
  {2012})%
  \bibAnnoteFile{NoStop}{NiJMB12}%
\bibitem{GibbonsBJ13}%
  \BibitemOpen
  \bibfield{author}{%
  \bibinfo {author} {\bibfnamefont{M.~M.}\ \bibnamefont{Gibbons}}\ and\
  \bibinfo {author} {\bibfnamefont{W.~S.}\ \bibnamefont{Klug}},\ }%
  \bibfield{journal}{%
  \bibinfo {journal} {Biophys.\ J.}\ }%
  \textbf{\bibinfo {volume} {95}},\ \bibinfo {pages} {3640} (\bibinfo {year}
  {2008})%
  \bibAnnoteFile{NoStop}{GibbonsBJ13}%
\bibitem{KlugPRL12}%
  \BibitemOpen
  \bibfield{author}{%
  \bibinfo {author} {\bibfnamefont{W.~S.}\ \bibnamefont{Klug}}, \bibinfo
  {author} {\bibfnamefont{W.~H.}\ \bibnamefont{Roos}},\ and\ \bibinfo {author}
  {\bibfnamefont{G.~J.~L.}\ \bibnamefont{Wuite}},\ }%
  \bibfield{journal}{%
  \bibinfo {journal} {Phys.\ Rev.\ Lett.}\ }%
  \textbf{\bibinfo {volume} {109}},\ \bibinfo {pages} {168104} (\bibinfo {year}
  {2012})%
  \bibAnnoteFile{NoStop}{KlugPRL12}%
\bibitem{TamaJMB05}%
  \BibitemOpen
  \bibfield{author}{%
  \bibinfo {author} {\bibfnamefont{F.}~\bibnamefont{Tama}}\ and\ \bibinfo
  {author} {\bibfnamefont{C.~L.}\ \bibnamefont{{Brooks 3rd}}},\ }%
  \bibfield{journal}{%
  \bibinfo {journal} {J.\ Mol.\ Biol.}\ }%
  \textbf{\bibinfo {volume} {345}},\ \bibinfo {pages} {299} (\bibinfo {year}
  {2005})%
  \bibAnnoteFile{NoStop}{TamaJMB05}%
\bibitem{YangBJ09}%
  \BibitemOpen
  \bibfield{author}{%
  \bibinfo {author} {\bibfnamefont{Z.}~\bibnamefont{Yang}}, \bibinfo {author}
  {\bibfnamefont{I.}~\bibnamefont{Bahar}},\ and\ \bibinfo {author}
  {\bibfnamefont{M.}~\bibnamefont{Widom}},\ }%
  \bibfield{journal}{%
  \bibinfo {journal} {Biophys.\ J.}\ }%
  \textbf{\bibinfo {volume} {64}},\ \bibinfo {pages} {4438} (\bibinfo {year}
  {2009})%
  \bibAnnoteFile{NoStop}{YangBJ09}%
\bibitem{ZinkBJ10}%
  \BibitemOpen
  \bibfield{author}{%
  \bibinfo {author} {\bibfnamefont{M.}~\bibnamefont{Zink}}\ and\ \bibinfo
  {author} {\bibfnamefont{H.}~\bibnamefont{Grubmuller}},\ }%
  \bibfield{journal}{%
  \bibinfo {journal} {Biophys.\ J.}\ }%
  \textbf{\bibinfo {volume} {98}},\ \bibinfo {pages} {687} (\bibinfo {year}
  {2010})%
  \bibAnnoteFile{NoStop}{ZinkBJ10}%
\bibitem{ArkhipovBJ09}%
  \BibitemOpen
  \bibfield{author}{%
  \bibinfo {author} {\bibfnamefont{A.}~\bibnamefont{Arkhipov}}, \bibinfo
  {author} {\bibfnamefont{W.~H.}\ \bibnamefont{Roos}}, \bibinfo {author}
  {\bibfnamefont{G.~J.~L.}\ \bibnamefont{Wuite}},\ and\ \bibinfo {author}
  {\bibfnamefont{K.}~\bibnamefont{Schulten}},\ }%
  \bibfield{journal}{%
  \bibinfo {journal} {Biophys.\ J.}\ }%
  \textbf{\bibinfo {volume} {97}},\ \bibinfo {pages} {2061} (\bibinfo {year}
  {2009})%
  \bibAnnoteFile{NoStop}{ArkhipovBJ09}%
\bibitem{CieplakJCP10}%
  \BibitemOpen
  \bibfield{author}{%
  \bibinfo {author} {\bibfnamefont{M.}~\bibnamefont{Cieplak}}\ and\ \bibinfo
  {author} {\bibfnamefont{M.~O.}\ \bibnamefont{Robbins}},\ }%
  \bibfield{journal}{%
  \bibinfo {journal} {J.\ Chem.\ Phys.}\ }%
  \textbf{\bibinfo {volume} {132}},\ \bibinfo {pages} {015101} (\bibinfo {year}
  {2010})%
  \bibAnnoteFile{NoStop}{CieplakJCP10}%
\bibitem{MayBJ11}%
  \BibitemOpen
  \bibfield{author}{%
  \bibinfo {author} {\bibfnamefont{E.~R.}\ \bibnamefont{May}}, \bibinfo
  {author} {\bibfnamefont{A.}~\bibnamefont{Aggarwal}}, \bibinfo {author}
  {\bibfnamefont{W.~S.}\ \bibnamefont{Klug}},\ and\ \bibinfo {author}
  {\bibfnamefont{C.~L.~B.}\ \bibnamefont{3rd}},\ }%
  \bibfield{journal}{%
  \bibinfo {journal} {Biophys.\ J.}\ }%
  \textbf{\bibinfo {volume} {100}},\ \bibinfo {pages} {L59} (\bibinfo {year}
  {2011})%
  \bibAnnoteFile{NoStop}{MayBJ11}%
\bibitem{KononovaBJ13}%
  \BibitemOpen
  \bibfield{author}{%
  \bibinfo {author} {\bibfnamefont{O.}~\bibnamefont{Kononova}}, \bibinfo
  {author} {\bibfnamefont{J.}~\bibnamefont{Snijder}}, \bibinfo {author}
  {\bibfnamefont{M.}~\bibnamefont{Brasch}}, \bibinfo {author}
  {\bibfnamefont{J.~J. L.~M.}\ \bibnamefont{Cornelissen}}, \bibinfo {author}
  {\bibfnamefont{R.~I.}\ \bibnamefont{Dima}}, \bibinfo {author}
  {\bibfnamefont{K.~A.}\ \bibnamefont{Marx}}, \bibinfo {author}
  {\bibfnamefont{G.~J.~L.}\ \bibnamefont{Wuite}}, \bibinfo {author}
  {\bibfnamefont{W.~H.}\ \bibnamefont{Roos}},\ and\ \bibinfo {author}
  {\bibfnamefont{V.}~\bibnamefont{Barsegov}},\ }%
  \bibfield{journal}{%
  \bibinfo {journal} {Biophys.\ J.}\ }%
  \textbf{\bibinfo {volume} {105}},\ \bibinfo {pages} {1893} (\bibinfo {year}
  {2013})%
  \bibAnnoteFile{NoStop}{KononovaBJ13}%
\bibitem{SnijderMicron12}%
  \BibitemOpen
  \bibfield{author}{%
  \bibinfo {author} {\bibfnamefont{J.}~\bibnamefont{Snijder}}, \bibinfo
  {author} {\bibfnamefont{I.~L.}\ \bibnamefont{Ivanovska}}, \bibinfo {author}
  {\bibfnamefont{M.}~\bibnamefont{Baclayon}}, \bibinfo {author}
  {\bibfnamefont{W.~H.}\ \bibnamefont{Roos}},\ and\ \bibinfo {author}
  {\bibfnamefont{G.~J.~L.}\ \bibnamefont{Wuite}},\ }%
  \bibfield{journal}{%
  \bibinfo {journal} {Micron}\ }%
  \textbf{\bibinfo {volume} {43}},\ \bibinfo {pages} {1343} (\bibinfo {year}
  {2012})%
  \bibAnnoteFile{NoStop}{SnijderMicron12}%
\bibitem{Landau}%
  \BibitemOpen
  \bibfield{author}{%
  \bibinfo {author} {\bibfnamefont{L.~D.}\ \bibnamefont{Landau}}\ and\ \bibinfo
  {author} {\bibfnamefont{E.~M.}\ \bibnamefont{Lifshitz}},\ }%
  \emph{\bibinfo {title} {Theory of Elasticity}}\ (\bibinfo {publisher}
  {Pergamon Press},\ \bibinfo {year} {1986})%
  \bibAnnoteFile{NoStop}{Landau}%
\bibitem{Johnson}%
  \BibitemOpen
  \bibfield{author}{%
  \bibinfo {author} {\bibfnamefont{K.~L.}\ \bibnamefont{Johnson}},\ }%
  \emph{\bibinfo {title} {Contact Mechanics}}\ (\bibinfo {publisher} {Cambridge
  University Press},\ \bibinfo {year} {1985})%
  \bibAnnoteFile{NoStop}{Johnson}%
\bibitem{Timoshenko}%
  \BibitemOpen
  \bibfield{author}{%
  \bibinfo {author} {\bibfnamefont{S.~P.}\ \bibnamefont{Timoshenko}},\ }%
  \emph{\bibinfo {title} {Theory of Elastic Stability}}\ (\bibinfo {publisher}
  {McGraw-Hill Book Company, Inc},\ \bibinfo {year} {1961})%
  \bibAnnoteFile{NoStop}{Timoshenko}%
\bibitem{Gumbel}%
  \BibitemOpen
  \bibfield{author}{%
  \bibinfo {author} {\bibfnamefont{E.~J.}\ \bibnamefont{Gumbel}},\ }%
  \emph{\bibinfo {title} {Statistics of Extremes}}\ (\bibinfo {publisher}
  {Dover Publications},\ \bibinfo {year} {2004})%
  \bibAnnoteFile{NoStop}{Gumbel}%
\bibitem{GibbonsPRE07}%
  \BibitemOpen
  \bibfield{author}{%
  \bibinfo {author} {\bibfnamefont{M.~M.}\ \bibnamefont{Gibbons}}\ and\
  \bibinfo {author} {\bibfnamefont{W.~S.}\ \bibnamefont{Klug}},\ }%
  \bibfield{journal}{%
  \bibinfo {journal} {Phys.\ Rev.\ E}\ }%
  \textbf{\bibinfo {volume} {75}},\ \bibinfo {pages} {031901} (\bibinfo {year}
  {2007})%
  \bibAnnoteFile{NoStop}{GibbonsPRE07}%
\bibitem{ZhmurovProt10}%
  \BibitemOpen
  \bibfield{author}{%
  \bibinfo {author} {\bibfnamefont{A.}~\bibnamefont{Zhmurov}}, \bibinfo
  {author} {\bibfnamefont{R.~I.}\ \bibnamefont{Dima}}, \bibinfo {author}
  {\bibfnamefont{Y.}~\bibnamefont{Kholodov}},\ and\ \bibinfo {author}
  {\bibfnamefont{V.}~\bibnamefont{Barsegov}},\ }%
  \bibfield{journal}{%
  \bibinfo {journal} {Proteins}\ }%
  \textbf{\bibinfo {volume} {78}},\ \bibinfo {pages} {2984} (\bibinfo {year}
  {2010})%
  \bibAnnoteFile{NoStop}{ZhmurovProt10}%
\bibitem{ZhmurovJPC11}%
  \BibitemOpen
  \bibfield{author}{%
  \bibinfo {author} {\bibfnamefont{A.}~\bibnamefont{Zhmurov}}, \bibinfo
  {author} {\bibfnamefont{K.}~\bibnamefont{Rybnikov}}, \bibinfo {author}
  {\bibfnamefont{Y.}~\bibnamefont{Kholodov}},\ and\ \bibinfo {author}
  {\bibfnamefont{V.}~\bibnamefont{Barsegov}},\ }%
  \bibfield{journal}{%
  \bibinfo {journal} {J.\ Phys.\ Chem.}\ }%
  \textbf{\bibinfo {volume} {115}},\ \bibinfo {pages} {5278} (\bibinfo {year}
  {2011})%
  \bibAnnoteFile{NoStop}{ZhmurovJPC11}%
\bibitem{DePabloPRL03}%
  \BibitemOpen
  \bibfield{author}{%
  \bibinfo {author} {\bibfnamefont{P.~J.}~\bibnamefont{de Pablo}}, \bibinfo
  {author} {\bibfnamefont{I.~A.~T.}~\bibnamefont{Schaap}}, \bibinfo {author}
  {\bibfnamefont{F.~C.}~\bibnamefont{MacKintosh}},\ and\ \bibinfo {author}
  {\bibfnamefont{C.~F.}~\bibnamefont{Schmidt}},\ }%
  \bibfield{journal}{%
  \bibinfo {journal} {Phys.\ Rev.\ Lett.}\ }%
  \textbf{\bibinfo {volume} {91}},\ \bibinfo {pages} {098101} (\bibinfo {year}
  {2003})%
  \bibAnnoteFile{NoStop}{DePabloPRL03}%
\bibitem{KononovaJACS14}%
  \BibitemOpen
  \bibfield{author}{%
  \bibinfo {author} {\bibfnamefont{O.}~\bibnamefont{Kononova}}, \bibinfo
  {author} {\bibfnamefont{Y.}~\bibnamefont{Kholodov}}, \bibinfo {author}
  {\bibfnamefont{K.~E.}~\bibnamefont{Theisen}}, \bibinfo {author}
  {\bibfnamefont{K.~A.}~\bibnamefont{Marx}}, \bibinfo {author}
  {\bibfnamefont{R.~I.}~\bibnamefont{Dima}}, \bibinfo {author}
  {\bibfnamefont{F.~I.}~\bibnamefont{Ataullakhanov}}, \bibinfo {author}
  {\bibfnamefont{E.~L.}~\bibnamefont{Grishchuk}},\ and\ \bibinfo {author}
  {\bibfnamefont{V.}~\bibnamefont{Barsegov}},\ }%
  \bibfield{journal}{%
  \bibinfo {journal} {J.\ Am.\ Chem.\ Soc.}\ }%
  \textbf{\bibinfo {volume} {136}},\ \bibinfo {pages} {17036} (\bibinfo {year}
  {2014})%
  \bibAnnoteFile{NoStop}{KononovaJACS14}%
\end{thebibliography}

\providecommand{\noopsort}[1]{}\providecommand{\singleletter}[1]{#1}%

\newpage
\begin{table*}
\caption{\label{tab:table1} Biomechanical and statistical characteristics determining the deformation and collapse of biological particles --- CCMV, TrV, and AdV: the Young's moduli for Hertzian deformation $E_H$ and beam-bending $E_b$, the force scale parameter $F_b^*$ and shape parameter $m$. The first (second) entries correspond to the exact (approximate) methods of parameter estimation. The model predictions for $F^{col}$ are compared with the peak forces (in parenthesis) from the spectra (Figs.~\ref{fig:fig2},~\ref{fig:fig3}). For TrV and AdV, the shell thickness was estimated in SM.}
\begin{ruledtabular}
\begin{tabular}{c c c c c c c}
System & $E_H$, GPa & $E_b$, GPa & $F_b^*$, nN & $m$ & $F^{col}$, nN \\
\hline
CCMV (2-fold symmetry; {\em in silico}) & 0.013/0.012 & 0.50/0.50 & 1.70/1.25 & 1.7/1.5 & 0.67/0.69 (0.68) \\
CCMV (quasi-2-fold symmetry; {\em in silico}) & 0.011/0.011 & 0.37/0.35 & 1.50/1.25 & 1.4/1.6 & 0.58/0.64 (0.68) \\
CCMV (quasi-3-fold symmetry; {\em in silico}) & 0.012/0.012 & 0.52/0.46 & 1.75/1.33 & 1.4/1.6 & 0.58/0.64 (0.68) \\
empty CCMV (average; {\em in vitro}) & 0.019/0.023 & 0.85/0.81 & 1.90/1.00 & 1.2/1.3 & 0.56/0.78 (0.71) \\
empty TrV (average; {\em in vitro}) & 0.030/0.036 & 0.94/0.81 & 1.90/1.1 & 1.1/1.2 & 0.70/1.02 (0.69) \\
full TrV (average; {\em in vitro}) & 0.140/0.140 & 0.95/0.84 & 8.10/5.5 & 1.1/1.0 & 2.91/3.78 (3.00) \\
full AdV (2-fold symmetry; {\em in vitro}) & 0.037/0.040 & 0.35/0.29 & 10.0/5.0 & 1.2/1.4 & 2.58/4.05 (3.80) \\
full AdV (3-fold symmetry; {\em in vitro}) & 0.018/0.019 & 0.20/0.18 & 11.0/5.0 & 1.3/1.7 & 3.04/4.15 (4.30)\\
full AdV (5-fold symmetry; {\em in vitro}) & 0.021/0.023 & 0.14/0.13 & 5.10/3.7 & 1.1/1.0 & 2.03/2.35 (1.90) \\
\end{tabular}
\end{ruledtabular}
\end{table*}
\end{document}



\title{Supplementary Material for ``Fluctuating nonlinear spring model of mechanical deformation of biological particles''}


\author{Olga Kononova}
\affiliation{Department of Chemistry, University of Massachusetts, Lowell, MA 01854, USA}
\affiliation{Moscow Institute of Physics and Technology, Moscow Region, 141700, Russia}

\author{Joost Snijder}
\affiliation{Natuur- en Sterrenkunde and LaserLab, Vrije Universiteit, 1081 HV Amsterdam, The Netherlands}

\author{Kenneth A. Marx}
\affiliation{Department of Chemistry, University of Massachusetts, Lowell, MA 01854, USA}

\author{Gijs J. L. Wuite}
\affiliation{Natuur- en Sterrenkunde and LaserLab, Vrije Universiteit, 1081 HV Amsterdam, The Netherlands}

\author{Wouter H. Roos}
\email[Corresponding author: ]{wroos@few.vu.nl}
\affiliation{Natuur- en Sterrenkunde and LaserLab, Vrije Universiteit, 1081 HV Amsterdam, The Netherlands}

\author{Valeri Barsegov}
\email[Corresponding author: ]{Valeri\_Barsegov@uml.edu}
\affiliation{Department of Chemistry, University of Massachusetts, Lowell, MA 01854, USA}
\affiliation{Moscow Institute of Physics and Technology, Moscow Region, 141700, Russia}


\date{\today}


\pacs{87.10.Ca,87.10.Pq}

\maketitle

\subsection{1. Self Organized Polymer (SOP) model of a virus particle\label{}}
The SOP model of the polypeptide chain was originally designed to address the mechanical properties of proteins \cite{HyeonStr06,MicklerPNAS07,ZhmurovProt10,ZhmurovJPC11}. The model has been applied to a variety of biological systems \cite{ZhangStr12,HyeonPNAS12,ZhmurovStr11,KononovaBJ13}. In this work, the SOP model has been used to describe each protein subunit forming a virus capsid. In the topology-based SOP model (Fig.~\ref{fig:figs2}), each amino acid residue is represented by a single interaction center described by the $C_\alpha$-atom, and the protein backbone is represented by a collection of the $C_\alpha$-$C_\alpha$ covalent bonds with the peptide bond length distance of $a = 3.8$ \AA. The potential energy function $U_{SOP}$ specified in terms of the coordinates of the $C_\alpha$-atoms $\{r_i\} = r_1, r_2,\dots, r_N$ ($N$ is the total number of residues) is given by:
\begin{equation} \label{eq:usop}
U_{SOP} = U_{FENE} + U_{NB}^{ATT} + U_{NB}^{REP}
\end{equation}
In Eq. (\ref{eq:usop}), the first term is the finite extensible nonlinear elastic (FENE) potential:
\begin{equation} \label{eq:ufene}
U_{FENE}=-\sum_{i=1}^{N-1}\frac{k R_0}{2}\log \left(1-\frac{(r_{i,i+1}-r_{i,i+1}^0)^2}{R_0^2}\right)
\end{equation}
where $k = 14$ N/m is the spring constant, and the tolerance in the change of the covalent bond distance is $R_0 = 2$ \AA. The FENE potential describes the backbone chain connectivity. The distance between the next-neighbor residues $i$ and $i+1$, is $r_{i,i+1}$, and $r^0_{i,i+1}$ is its value in the native  structure. To account for the non-covalent (non-bonded) interactions that stabilize the native state, we use the Lennard-Jones potential:
\begin{equation} \label{eq:uatt}
U_{NB}^{ATT} = \sum_{i,j=i+3}^{N-3}\varepsilon_h\left[ \left( \frac{r_{ij}^0}{r_{ij}} \right)^{12} - 2\left( \frac{r_{ij}^0}{r_{ij}} \right)^{6} \right]\Delta_{ij}
\end{equation}
In Eq. (\ref{eq:uatt}), we assume that if the non-covalently linked residues $i$ and $j$ ($|i-j| > 2$) are within the cut-off distance of $8$ \AA~in the native state, then $\Delta_{ij} = 1$; $\Delta_{ij} = 0$ otherwise. The value of $\varepsilon_h$ quantifies the strength of the non-bonded interactions. The non-native (non-bonded) interactions are treated as repulsive:
\begin{equation} \label{eq:urep}
U_{NB}^{REP} = \sum_{i,j=i+2}^{N-2}\varepsilon_r \left(\frac{\sigma_r}{r_{ij}} \right)^6 + \sum_{i,j+i+3}^{N-3}\varepsilon_r \left( \frac{\sigma_r}{r_{ij}} \right)^6(1-\Delta_{ij})
\end{equation}
In Eq. (\ref{eq:urep}), a constraint is imposed on the bond angle between the residues $i$, $i+1$, and $i+2$ by including the repulsive potential with parameters $\varepsilon_l = 1$ kcal/mol and $\sigma_l = 3.8$ \AA. These define the strength and the range of the repulsion. In the SOP model, parameter εh sets the energy scale. This parameter is estimated based on the results of all-atom MD simulations of the virus particle at equilibrium. \\
\indent The dynamics of the virus system is obtained by solving numerically the Langevin equations of motion for each particle position $r_i$ in the over-damped limit:
\begin{equation} \label{eq:ld}
\eta \frac{dr_i}{dt} = - \frac{\partial U_i(r_i)}{\partial r_i} + g_i(t)
\end{equation}
In Eq. (\ref{eq:ld}), $U_i(r_i)$ is the total potential energy, which accounts for the biomolecular interactions ($U_{SOP}$) and interactions of particles with the indenting object --- spherical tip ($U_{tip}$; see Eq. (\ref{eq:utip}) in section 3 below). Also, in Eq. (\ref{eq:ld}) $g_i(t)$ is the Gaussian distributed zero-average random force, and $\eta$ is the friction coefficient. To generate the Brownian dynamics, the equations of motion for each $C_\alpha$-atom are propagated with the time step $\Delta t = 0.08\tau_H$, where $\tau_H = \zeta \varepsilon_h \tau_L/k_BT$ ($\Delta t = 20$ ps for CCMV). Here, $\tau_L = (m a^2/\varepsilon_h)^{1/2} = 3$ ps, $\zeta = 50.0$ is the dimensionless friction constant for an amino acid residue in water ($\eta = \zeta m/ \tau_L$ ), $m \approx 3 \times 10^{22}$ g is the residue mass, and $T$ is the absolute temperature \cite{BarsegovBJ06,ZhmurovBJ10}. To perform simulations of nanoindentation of a virus particle, we set $T$ to room temperature and use the bulk water viscosity, which corresponds to the friction coefficient $\eta = 7.0 \times 10^5$ pN ps/nm.

\subsection{2. SOP model parameterization for CCMV shell}
\begin{figure}
\includegraphics[width=\textwidth]{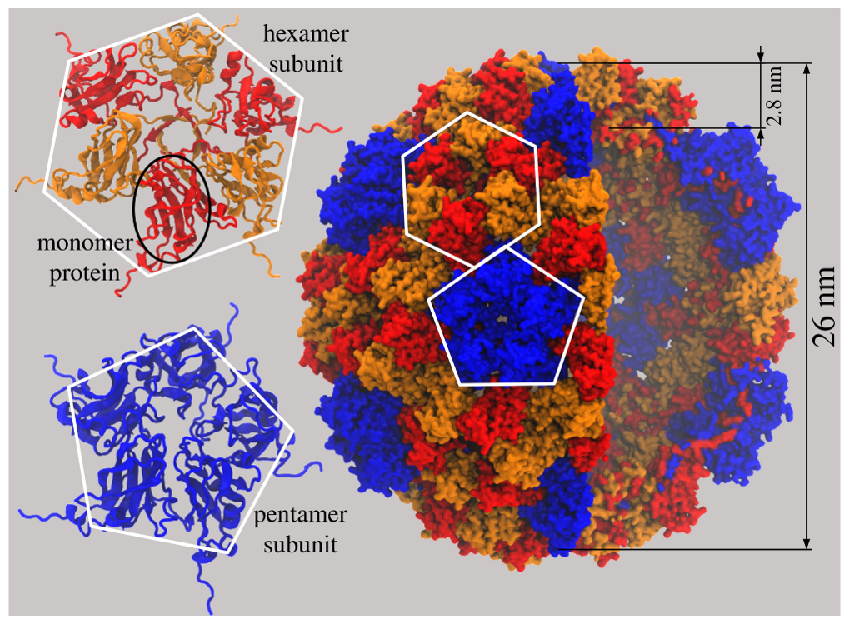}%
\caption{\label{fig:figs1} The structure of the Cowpea Chlorotic Mottle Virus (CCMV) (PDB code: 1CWP). The side view of the CCMV shell is shown on the right. The protein domains forming pentamers are in blue, while the same protein domains in hexamers are in red and orange. The hexamers and pentamers, composed of six and five copies of the same protein chain (circled in the black ellipse), are displayed on the left. The CCMV capsid is a $\sim$$2.8$ nm thick shell with a $\sim$$26$ nm diameter.}
\end{figure}

\begin{figure}
\includegraphics[width=0.6\textwidth]{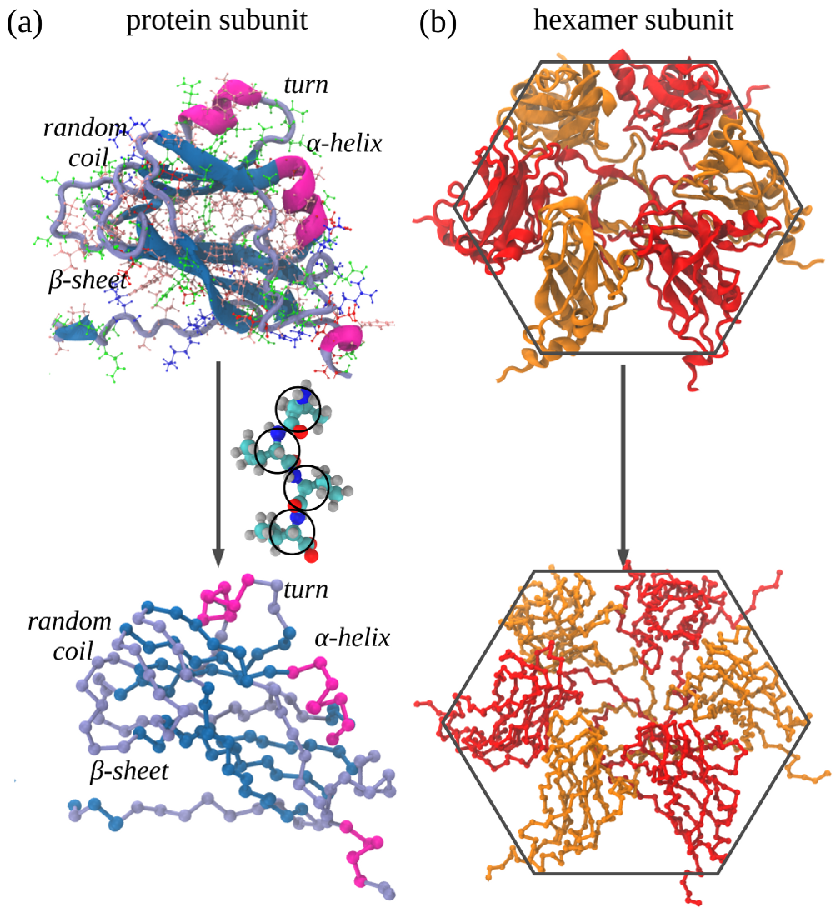}%
\caption{\label{fig:figs2} Graphical illustration of the coarse-graining procedure involved in construction of a SOP model \cite{HyeonStr06,MicklerPNAS07} of a polypeptide chain (sections 1 and 2 in SM). Panel (b) shows coarse-graining of the atomic structure of the protein subunit forming pentamers and capsomers of the CCMV shell (Fig. S\ref{fig:figs1}). Each amino acid residue is represented by a spherical bead of an appropriate radius with the coordinates of the $C_\alpha$-atom (black circles). The protein backbone is replaced by a collection of the $C_\alpha$-$C_\alpha$ covalent bonds with $3.8$ \AA~bond distance. The potential energy function (see Eq. (\ref{eq:usop})) describes the interactions between amino acids stabilizing the native state of the protein chain, and the chain connectivity, elongation due to stretching, and self-avoidance. The coarse-graining procedure preserves the secondary structure: $\alpha$-helices (pink), $\beta$-strands and sheets (blue), and random coil and turns (gray). Panel (b) shows the results of coarse-graining of a hexamer. Six identical copies of the same protein monomer (coarse-grained in (a)) form a Cα-based model of the hexamer subunit. The hexamers and pentamers are combined to form a coarse-grained reconstruction of the full CCMV shell. The SOP model describes the geometry and 3D shape of the biological particle.}
\end{figure}

For each virus system, the numerical value of the parameter $\varepsilon_h$, which describes the strength of native interactions (section 1), has been determined from the all-atom MD simulations. We performed the all-atom MD simulations of the atomic structural model of the CCMV shell (Fig.~\ref{fig:figs1}). To obtain accurate parameterization of the SOP model for CCMV, we used the crystal structure of the capsid (Viper Data Base; PDB code: 1CWP \cite{SpeirStr95} with $T = 3$ symmetry) in conjunction with the Solvent Accessible Surface Area (SASA) model of implicit solvation (CHARMM19 force field) \cite{FerraraProt02}. First, we calculated the number of native contacts using a standard cut-off distance between the $C_\alpha$-atoms of $8.0$ \AA. The native contacts were divided into the inter-chain contacts and the intra-chain contacts. For the CCMV shell, there are $N_{intra} \approx 20,554$ intra-chain contacts that stabilize the native state of the capsid protein, and $N_{inter} \approx 3,405$ inter-chain contacts at the interfaces formed by capsid proteins. Next, we calculated the total energy of non-covalent interactions for each contact group. The total energy for the intra-chain contacts is $E_{intra} = 25,898$ kcal/mol and the total energy for the inter-chain contacts is $E_{inter} = 3,746$ kcal/mol. Finally, to obtain the values of parameter $\varepsilon_h$ we divided the two numbers for each contact group. For the CCMV shell, we obtained $\varepsilon_{intra} = E_{intra}/N_{intra} = 1.26$ kcal/mol and $\varepsilon_{inter} = E_{inter}/N_{inter} = 1.1$ kcal/mol. The atomic-level details of biomolecular interactions, contained in these parameters, were then exported into the SOP model based reconstruction of the full CCMV particle (Fig.~\ref{fig:figs2}).

\subsection{3. Nanoindentation \textit{in silico} method}
\begin{figure}
\includegraphics[width=0.75\textwidth]{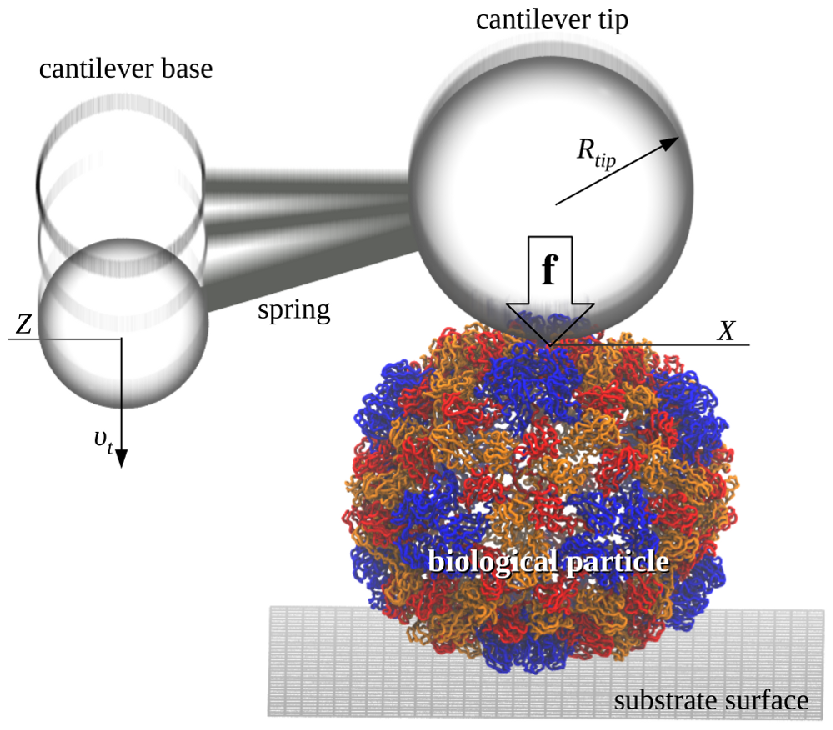}%
\caption{\label{fig:figs3} Schematic of the setup used in nanoindentations \textit{in vitro} or \textit{in silico} (see section 3 in SM). The biological particle (virus shell) is placed on the substrate. The cantilever base (virtual sphere) is moving in the direction perpendicular to the surface of the particle with the constant velocity $\nu_f$ (force-ramp), which creates a compressive force. The force is transmitted to the cantilever tip (sphere of radius $R_{tip}$) through the harmonic spring with the spring constant $\kappa$. The force exerted on a particle $f(t) = r_ft$ (large vertical arrow) ramps up linearly in magnitude with time with the force-loading rate $r_f = \kappa \nu_f$, which mechanically loads the particle. The mechanical response of the particle can be probed by profiling the deformation force (indentation force) $F$ as a function of the cantilever base (piezo-) displacement $Z$ ($FZ$ curve) or as a function of the indentation depth $X$ ($FX$ curve).}
\end{figure}

We employed the methodology of “nanoindentation \textit{in silico}” (Fig.~\ref{fig:figs3}) \cite{KononovaBJ13}, in which the mechanical loading of a biological particle is performed computationally. The nanoindentation measurements are performed under experimental conditions of dynamic force application $f(t) = r_f t$ (Fig.~\ref{fig:figs3}), i.e. in our simulations we use the experimental force-loading rates $r_f$, which correspond to the cantilever base velocity $\nu_f = 0.1 - 1.0$ $\mu$m/s. Structural transitions can be resolved by examining the coordinates of amino acid residues, and biomechanical characteristics can be gathered through analysis of the energy output. Because our \textit{in silico} ``experiment'' provides a complete simulation view of particle deformation and collapse, it can be used to provide a detailed interpretation and modeling of the experimental force-deformation spectra. The full control we have over the system during the nanomanipulations \textit{in silico can} be used to study deformation at different symmetry points on the particle surface, and to relate the force and energy values recorded at any point in the simulation to the specific details observed in the particle's structure. \\
\indent In dynamic force measurements \textit{in silico}, the cantilever base is represented by the virtual particle, connected to the spherical bead of radius $R_{tip}$, mimicking the cantilever tip (indenter), by a harmonic spring (Fig.~\ref{fig:figs3}). The tip interacts with the particles via the repulsive Lennard-Jones potential
\begin{equation} \label{eq:utip}
U_{tip} = \sum_{i=1}^{N}{\varepsilon_{tip} \left( \frac{\sigma_{tip}}{|r_i - r_{tip} - R_{tip}} \right)^6}
\end{equation}
thereby producing an indentation on the particle's outer surface. In Eq. (\ref{eq:utip}), $r_i$ and $r_{tip}$ are coordinates of the ith particle and the center of the tip, respectively, $\varepsilon_{tip} = 4.18$ kJ/mol, and $\sigma_{tip} = 1.0$ \AA are parameters of interaction, and the summation is performed over all the particles under the tip. For the cantilever tip (sphere in Fig.~\ref{fig:figs3}), we solve numerically the following Langevin equation of motion:
\begin{equation} \label{eq:ldtip}
\eta \frac{dr_{tip}}{dt} = - \frac{\partial U_{tip}(r_{tip})}{\partial r_{tip}} + \kappa((r_{tip}^0 - \nu_f t) - r_{tip})
\end{equation}
where $r_{tip}^0$ is the initial position of spherical tip center ($\nu_f$  is the cantilever base velocity; $\kappa$ is the cantilever spring constant), and the friction coefficient $\eta = 7.0 \times 10^6$ pN ps/nm. To generate the dynamics of the biological particle of interest tested mechanically, we solve numerically Eqs (\ref{eq:usop}) --- (\ref{eq:ld}) for the particle (see section 1) and Eqs. (\ref{eq:utip}) and (\ref{eq:ldtip}) for the indenter (spherical tip). \\
\indent The cantilever base moving with constant velocity ($\nu_f$) (Fig.~\ref{fig:figs3}) exerts (through the tip) the time-dependent force (force-ramp) ${\bf f}(t) = f(t){\bf n}$ in the direction ${\bf n}$ perpendicular to the particle surface. The force magnitude, $f(t) = r_f t$, exerted on the particle increases linearly in time $t$ with the force-loading rate $r_f = \kappa \nu_f$. In the simulations of ``forward indentation'', the cantilever base (and spherical tip) is moving towards the virus capsid. We control the piezo (cantilever base) displacement $Z$, and the cantilever tip position $X$, which defines the indentation depth (deformation). The resisting force of deformation $F$ from the virus particle, which corresponds to the experimentally measured force is calculated using the energy output. To prevent the capsid from rolling, we can always constrain the bottom portion of the particle by fixing select $C_\alpha$-atoms contacting the substrate surface.

\subsection{4. Method of Lagrange multipliers}
To study the dynamical changes in $x_H$ (Hertzian deformation) and $x_b$ (beam-bending deformation) and their contribution to the total deformation $X = x_H + x_b$ (see Fig. 2 in the main text), we employed the method of Lagrange multipliers \cite{McQuarrie}. This method allows us to find the values of $x_H$ and $x_b$ that minimize the total deformation force, $F(x_H,x_b) = k_H x_H^{3/2} + K_b x_b s(x_b)$ (see \textit{the inset} Fig. 3 in main text), subject to the constraint: $X = x_H + x_b$. To that end, we constructed the Lagrange function $\Lambda(x_H, x_b, \lambda) = F(x_H,x_b) + \lambda g(x_H,x_b)$, where $g(x_H,x_b) = x_H + x_b - X$ and $\lambda$ is the Lagrange multiplier. By calculating the partial derivatives of $\Lambda(x_H, x_b, \lambda$) with respect to each of the two variables $x_H$ and $x_b$, we obtained equations of the form $\nabla_{x_H,x_b}F(x_H,x_b) = -\lambda \nabla_{x_H,x_b}g(x_H,x_b)$. Next, by eliminating $\lambda$ we arrived at the system of two equations:
\begin{equation}\label{eq:lagrange}
3/2 k_H x_H^{1/2} = K_b s(x_b) + K_b x_b s^\prime(x_b) 
\end{equation}
\begin{equation}\label{eq:lconstr}
X = x_h + x_b
\end{equation}
For small deformations $x_b$, we expand the Weibull survival probability $s(x_b) = \exp[-(K_b x_b / F_b^*)^m]$ in powers of the exponent $z = K_b x_b / F_b^*$, and retain the terms up to the first order in $z$. Then, Eq. (~\ref{eq:lagrange}) becomes:
\begin{equation}\label{eq:lagrange2}
3/2 k_H x_H^{1/2} - K_b(1 - (K_b x_b/F_b^*)^m)(1 - m(K_b x_b/F_b^*)^m) = 0
\end{equation}
Eq. (~\ref{eq:lconstr}) allows us to eliminate $x_H$ by substituting $x_H = X - x_b$ into Eq. (\ref{eq:lagrange2}) above:
\begin{equation}\label{eq:lagrange3}
3/2 k_H (X - x_b)^{1/2} - K_b(1 - (K_b x_b/F_b^*)^m)(1 - m(K_b x_b/F_b^*)^m) = 0
\end{equation}
Simplifying Eq. (\ref{eq:lagrange3}) and grouping terms of the same power in $x_b$, we arrive at the following polynomial equation:
\begin{equation}\label{eq:lpolynom}
a_1x_b^{4m} + a_2x_b^{3m} + a_3x_b^{2m} + a_4x_b^{m} + a_5x_b + a_6 = 0
\end{equation}
which has the following constant coefficients: $a_1 = mK_b^2(K_b/F_b^*)^{4m}$, $a_2 = -2m(1+m)K_b^2(K_b/F_b^*)^{3m}$, $a_3 = (1+4m+m^2)K_b^2(K_b/F_b^*)^{2m}$, $a_4 = -2(1+m)K_b^2(K_b/F_b^*)^m$, $a_5 = 9/4k_H^2$, and $a_6 = K_b^2 - 9/4k_H^2X$. Eq. (S~\ref{eq:lpolynom}) can be solved numerically (for example, using Mathematica software) for a given set of parameters $k_H$, $K_b$, $F_b^*$, and $m$, and for each specified value of the total deformation $X$. The obtained numerical solution for $x_H$ and $x_b$ can then be substituted into the expression for $F(x_H, x_b)$ (Eq. (10) in the main text).

\subsection{5. Estimation of the thickness of TrV and AdV with encapsulated genome}
When a virus particle is empty (i.e. it does not contain genomic material) the thickness of the beams is equal to the virus shell thickness. The capsid thickness with encapsulated genome due to DNA/RNA packing can be estimated assuming that DNA/RNA molecules are evenly distributed on the inner capsid surface. The volume occupied by the genome considered to form a long cylindrical tube is given by $V_{gen} = \pi R_{gen}^2 L_{gen}$, where $L_{gen}$ is the genome length and $R_{gen} = 1.0$ nm and $0.5$ nm is the radius of cross-sectional area of the dsDNA molecule (for AdV) and ssRNA molecule (for TrV), respectively. The inner volume of the virus particle, $V_{in} = V_{gen} = V_{emp}$, is the sum of the volume occupied by the genome ($V_{gen}$) and the volume of the empty space ($V_{emp}$). Here, $V_{in} = 4/3 \pi R_{in}^3$ and $V_{emp} = 4/3 \pi (R_in - r_gen)^3$, where $R_{in} = R_{par} - r$ is the inner radius of the particle shell, $r$ is the shell thickness (see Eq. 2 in main text), and $r_{gen}$ is the increase in shell thickness due to encapsulated genome. This allows us to express rgen as $r_{gen} = R_{in} - (R_{in}^3 - 3/4 R_{gen}^2 L_{gen})^{1/3}$. For the AdV shell, $R_{in} = 36$ nm and $L_{gen} = 12$ $\mu$m \cite{SanMartinViruses12}; hence, $r_{gen} = 2.5$ nm. For the TrV shell, $R_{in} = 11$ nm and $L_{gen} = 3.06$ $\mu$m \cite{SnijderNC13}; hence, $r_{gen} = 1.9$ nm. These values of $r_{gen}$ for AdV and TrV shells were used to estimate $E_b$ (Table I in main text).

\providecommand{\noopsort}[1]{}\providecommand{\singleletter}[1]{#1}%

\newpage

\begin{figure}
\includegraphics[width=0.6\textwidth]{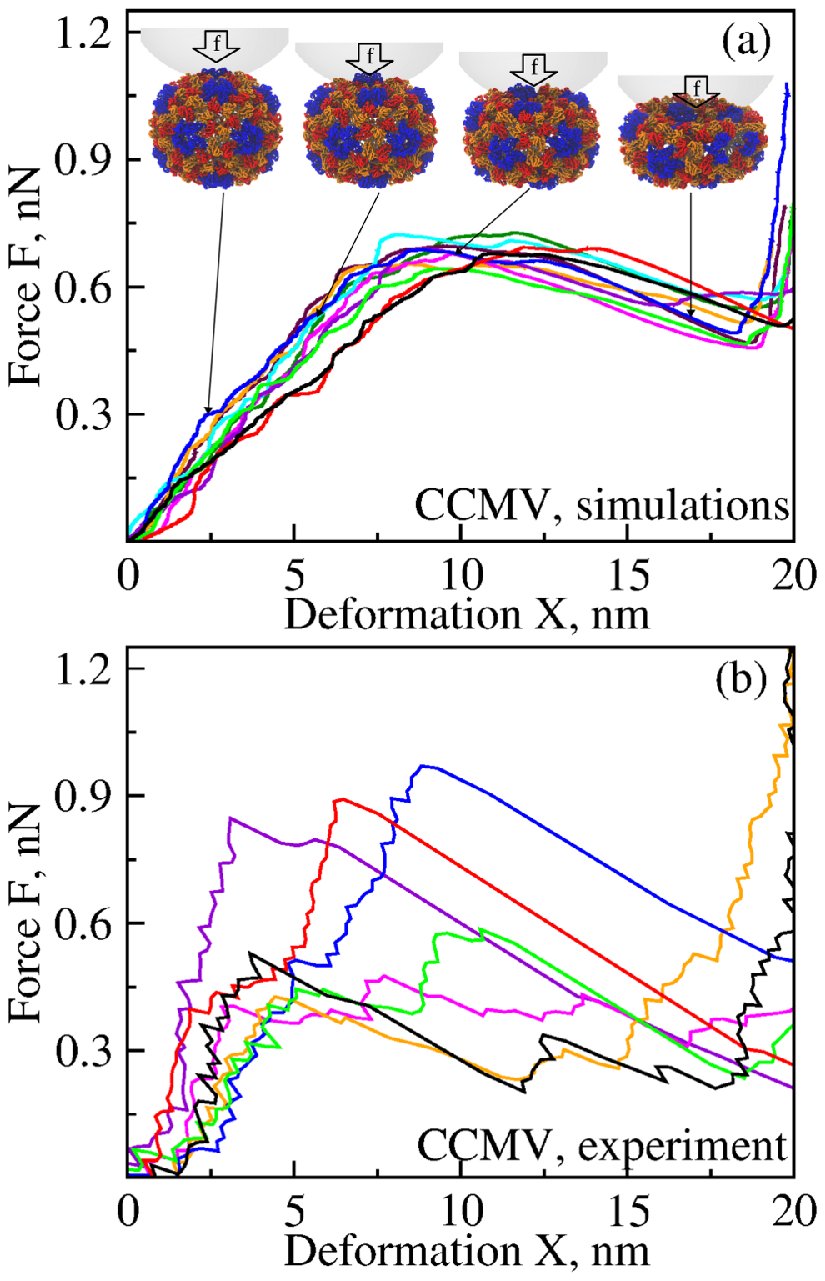}%
\caption{\label{fig:figs4} Nanoindentation of the empty CCMV particle \textit{in silico} (a) and \textit{in vitro} (b). Shown in different colors for clarity are the $FX$ curves obtained using the cantilever tip velocity $\nu_f = 0.06$ $\mu$m/s (experiment) and $\nu_f = 1.0$ $\mu$m/s (simulations). In the AFM-based experiments and in simulations of nanoindentation of CCMV, we used the cantilever tip with radius $R_{tip} = 20$ nm and the spring constant $\kappa = 0.05$ N/m. In panel (a), structural snapshots from the left to the right, which correspond to the $FX$ curve shown in blue, display the progress of forced deformation from the native un-deformed state (leftmost structure), to the partially deformed state (middle structures), and finally to the globally collapsed state (rightmost structure). In nanoindentation measurements \textit{in silico} and \textit{in vitro}, the cantilever tip indents the capsid in the direction perpendicular to the capsid outer surface (shown by a large vertical arrow).}
\end{figure}

\begin{figure}
\includegraphics[width=0.45\textwidth]{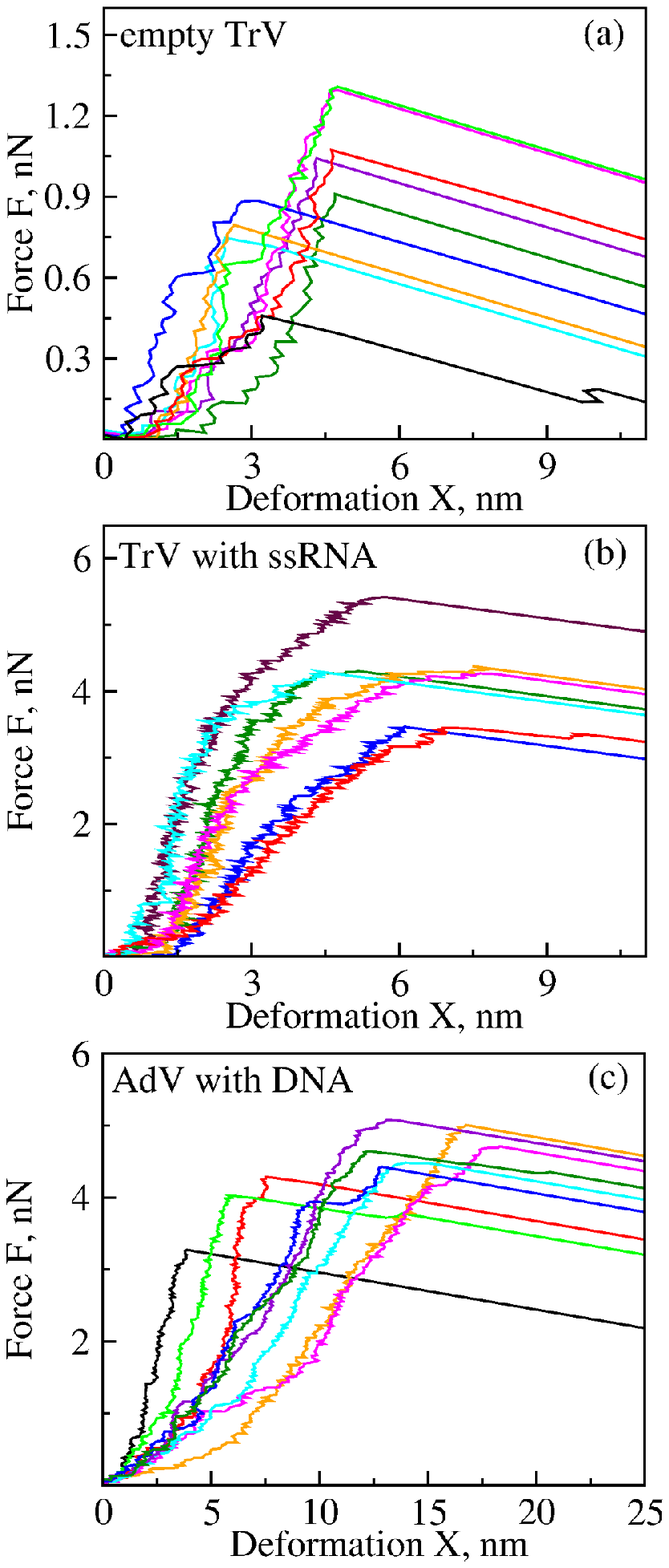}%
\caption{\label{fig:figs5} AFM-based nanoindentation of the empty TrV particle (a), full TrV particle (encapsulating the single-stranded RNA genome; (b)) and full AdV particle (encapsulating the DNA genome, (c)). Shown in different colors for clarity are the representative force-deformation spectra. The $FX$ curves for the empty TrV particle were obtained using the cantilever tip velocity $\nu_f = 0.06$ $\mu$m/s, tip radius $R_{tip} = 15$ nm, and spring constant $\kappa = 0.056$ N/m. The $FX$ curves for the full TrV particle were obtained using $\nu_f = 0.06$ $\mu$m/s, $R_{tip} = 15$ nm, and $\kappa = 0.1$ N/m. The $FX$ curves for the full AdV particle were obtained using $\nu_f = 0.055$ $\mu$m/s, $R_{tip} = 15$ nm, and $\kappa = 0.0524$ N/m.}
\end{figure}